%% file: main.tex
\documentclass[lettersize,journal]{IEEEtran}
\IEEEoverridecommandlockouts

\usepackage{amsmath,amsfonts,amssymb}
\usepackage{algorithmic}
\usepackage{algorithm}
\usepackage{array}
\usepackage[caption=false,font=normalsize,labelfont=sf,textfont=sf]{subfig}
\usepackage{textcomp}
\usepackage{stfloats}
\usepackage{url}
\usepackage{verbatim}
\usepackage{graphicx}
\usepackage{cite}
\usepackage{xcolor}
\usepackage{float}
\usepackage{tabularx}
\usepackage{booktabs}

\usepackage[font=small]{caption}
\usepackage{subfig}
\usepackage[acronym]{glossaries}
\input{acronyms}
\usepackage{tikz}
\usetikzlibrary{arrows.meta,positioning,shapes.geometric}
\usepackage[normalem]{ulem}
\usepackage{tikz}
\usepackage{pgfplots}
\usetikzlibrary{plotmarks}
\pgfplotsset{compat=1.18}
\usepackage{tikz}
\usetikzlibrary{positioning,calc,arrows.meta,decorations.pathmorphing}

\definecolor{bg}{RGB}{224,229,247}
\definecolor{frameblue}{RGB}{30,40,120}
\definecolor{linkgreen}{RGB}{0,160,80}
\definecolor{linkblue}{RGB}{70,90,210}
\definecolor{aiRed}{RGB}{220,0,0}
\definecolor{ricBlue}{RGB}{30,80,190}

\tikzset{
  >=Stealth,
  box/.style={draw=black, very thick, rounded corners=1.2pt, fill=white,
              minimum width=14mm, minimum height=9mm, align=center, font=\sffamily\small},
  sbox/.style={draw=black, thick, rounded corners=1pt, fill=white,
               minimum width=10mm, minimum height=7mm, align=center, font=\sffamily\scriptsize},
  roadm/.style={box, minimum width=15mm, minimum height=7mm, font=\sffamily\scriptsize},
  sw/.style={box, fill=gray!15, minimum width=17mm, minimum height=8mm, font=\sffamily\scriptsize},
  legend/.style={font=\sffamily\scriptsize},
  ai/.style={fill=aiRed, draw=aiRed, minimum width=5mm, minimum height=3.5mm},
  gLink/.style={draw=linkgreen, very thick},
  bLink/.style={draw=linkblue, very thick},
  sep/.style={draw=black, very thick, dashed},
}

\begin{document}

\title{Energy–Latency Trade-offs in O-RAN with Distributed Baseband Processing and AI Inference}

\markboth{Journal of \LaTeX\ Class Files,~Vol.~xx, No.~x, Month~YYYY}%
{Shell \MakeLowercase{\textit{et al.}}: A Sample Article Using IEEEtran.cls for IEEE Journals}

\author{
Urooj Tariq$^{*}$,
Rishu Raj$^{*}$,
Shashi Raj Pandey$^{\dagger}$,
Merim Dzaferagic$^{*}$,
Petar Popovski$^{\dagger}$,
Dan Kilper$^{*}$\\[4pt]

$^{*}$CONNECT Centre, Trinity College Dublin, Ireland\\
$^{\dagger}$Department of Electronic Systems, Aalborg University, Denmark\\
}

\maketitle

\begin{abstract}
The Open Radio Access Network (O-RAN) architecture introduces flexible functional splits and open interfaces that enable distributed and centralized deployment of baseband processing. While this flexibility offers opportunities for improved resource utilization, it also introduces fundamental trade-offs between energy efficiency and latency. In this paper, we develop a throughput-based end-to-end energy consumption model for O-RAN and extend it by incorporating detailed latency modeling and application-specific \gls{ai}/ \gls{ml} inference costs. The proposed end-to-end modeling framework provides a general representation of processing, transport, and inference-related energy and delay across the access, metro, and long-haul network segments. Building on this general model, we formulate an optimization problem that selects the placement of baseband processing and AI inference tasks across candidate O-RAN configurations to analyze energy-latency tradeoffs under network load, server frequency, and energy-budget constraints. Using representative hardware platforms and realistic traffic assumptions, we evaluate multiple baseband processing placements corresponding to different O-RAN functional configurations. Our results reveal how user quality of service requirements and network load conditions jointly determine the optimal placement of baseband processing and AI inference tasks, highlighting the inherent trade-off between energy efficiency and latency. The analysis provides practical insights for latency-aware and energy-efficient O-RAN deployments supporting emerging AI-driven services.
\end{abstract}


\section{Introduction}
\IEEEPARstart {T}{he} 5G standard sought to address heterogeneous performance requirements through the service categories \gls{embb}, \gls{urllc}, and \gls{mmtc}, but emerging applications often require combinations of these capabilities. As research moves towards sixth-generation (6G) systems, AI-native and highly virtualized architectures are gaining prominence. Within this evolution, the \gls{o-ran} specification \cite{ORANSpec95:online} has emerged as a key enabler, defining open and interoperable interfaces that support multi-vendor deployments and AI-driven control. Moreover, O-RAN-based infrastructures are increasingly expected to support AI-native services, including edge inference and generative workloads, which impose simultaneous demands on throughput, latency, and energy efficiency. Such applications influence network control and introduce application-level inference workloads whose energy and latency must be jointly considered with the underlying O-RAN transport and processing architecture. AI inference latency has been
widely studied, and practical acceleration techniques (e.g., quantization, pruning, and architecture-level optimizations) can reduce inference time under strict request-level latency budgets~\cite{zhang2021nn}. Due to O-RAN disaggregation, baseband processing and AI/ML tasks may be executed at different network nodes along the processing chain. The choice of where to execute these functions couples compute and transport costs, thereby creating inherent energy–latency trade-offs that must be evaluated jointly.

O-RAN operationalizes AI/ML through a hierarchical control architecture with well-defined timescales. Specifically, the Non-Real-Time RAN Intelligent Controller (Non-RT RIC) supports long-timescale control loops (typically seconds to minutes, i.e., $>1\,\mathrm{s}$) and hosts rApps for functions such as policy configuration, orchestration, and ML model training and lifecycle management. In contrast, the Near-Real-Time RIC (Near-RT RIC) enables monitoring and closed-loop optimization of O-CU/O-DU operations over near-real-time timescales (10 $\mathrm{ms}$ –1 $\mathrm{s}$), and hosts xApps to perform sub-second control tasks including policy enforcement and radio resource management. Real-time radio operations (i.e., $< 10 \,\mathrm{ms}$), including \gls{harq}, beamforming, and scheduling, are executed within the O-DU scheduler. This framework enables ML-assisted policies to be deployed and executed across the RAN control and orchestration stack\cite{garcia2021ran}.
 
Recent work shows that although ML-driven O-RAN architectures improve adaptability and resource efficiency, the energy footprint of AI model training, inference, and disaggregated cloud execution becomes a critical design factor, making accurate energy modeling of O-RAN components essential for sustainable deployment \cite{10552840}. Furthermore, energy is closely linked to the placement of baseband processing across O-RAN nodes, since different functional configurations shift the balance between compute power and transport power along the fronthaul/midhaul path\cite{11162430}, potentially increasing transport demand and tightening end-to-end latency. Beyond energy concerns, O-RAN must also meet stringent latency requirements over its transport network. In particular, the Open Fronthaul interface is highly time sensitive and can be vulnerable to delay and jitter when carried over shared transport networks \cite{10287312}. However, achieving energy-efficient O-RAN deployments that can still meet stringent requirements such as Enhanced Ultra Reliable Low Latency Communication (eURLLC) \cite{10054381} remains a critical challenge.

This work considers multiple RAN architectures, each assuming a different level of centralization of the baseband functions. These configuration choices lead to different power consumption, latency, and bandwidth requirements across radio and transport networks~\cite{8479363}. As illustrated in Fig.~\ref{fig:blockdiagram}, the considered AI-native O-RAN architecture consists of Open-Radio Unit (O-RU), Open-Distributed Unit (O-DU), Open-Centralised Unit (O-CU), and Data Center (DC) nodes connected through fronthaul, midhaul, and backhaul segments. The end-to-end energy consumption and latency arise from the joint behavior of baseband processing, transport, and AI/ML inference, rather than from a single component.

Different functional configuration options determine where the baseband functions are executed, thus changing the processing load at the radio site, edge, and cloud. In addition, AI/ML inference introduces application-level computation and memory-access latency, which contribute to end-to-end energy consumption and delay. Therefore, evaluating RAN architectures only from the perspective of network energy, communication latency, or AI inference cost provides an incomplete view of their suitability for AI-native deployments.

Although prior studies have examined energy consumption, latency, or AI inference costs in O-RAN systems, these aspects are typically analyzed in isolation. In contrast, we combine a throughput-based, end-to-end O-RAN energy consumption model with an application-oriented view of ML task inference latency, jointly capturing both network-level and ML-level contributions to overall energy use and delay. The goal is to understand which RAN architecture leads to the best energy performance while also satisfying latency requirements, thereby supporting future design choices for mobile network operators. To the best of our knowledge, this work is the first to combine throughput-based energy modeling, network latency analysis, and application-level AI inference costs into a single end-to-end framework for AI-native O-RAN deployments. Motivated by this gap, this paper makes the following contributions:

\begin{itemize}
    \item We develop an end-to-end model that decomposes energy and delay into baseband processing, transport, and AI inference components across O-RAN functional configurations.

    \item We extend throughput-based O-RAN energy models by incorporating application-level AI inference costs and fine-grained latency modeling, enabling a comprehensive energy–latency trade-off analysis under AI-native workloads.

    \item We formulate a latency-aware functional configuration optimization problem under energy constraints, providing a systematic method to select baseband and AI inference placement based on Quality of Service (QoS) requirements.

    \item Through extensive numerical evaluation, we characterize how network load, server frequency, and processing placement jointly determine feasibility regions for latency-critical services, revealing design guidelines for energy-efficient O-RAN deployments.
\end{itemize}

The remainder of the paper is organized as follows. Section II provides a detailed review of related research work. Section III presents the O-RAN architecture and describes the existing mathematical models used to evaluate the energy consumption and latency in O-RAN communication networks. Section IV presents our optimization formulation, describing the objective and key constraints. Section V reports and explains the results of our study, and Section VI concludes the paper.

\section{Related Work}
In this section, we outline the state-of-the-art research related to our work. With the advent of next-generation mobile networks, designing energy-efficient and flexible radio access infrastructures has become a central objective. In particular, there is a growing need to reduce network-wide power consumption while still satisfying the stringent requirements of Enhanced Ultra-Reliable Low-Latency Communication (eURLLC) services \cite{10054381}. A concise yet comprehensive overview of the O-RAN architecture, its standardized interfaces, typical power consumption models, and the role of machine learning for control and optimization in O-RAN is presented in \cite{10552840}. 
Furthermore, \cite{11162430} presents a throughput-based end-to-end power model that captures how traffic demand translates into energy consumption across both transport and processing elements in O-RAN, whereas \cite{10592054} adopts a parameterized formulation that focuses on component level power characterization, While this paper, we build on the throughput centric view in \cite{11162430} and extend it with additional latency and AI/ML task modeling.

Beyond modeling, a substantial body of work focuses on the energy-aware placement of baseband functions and functional split options in O-RAN. For example, \cite{9717286} formulates a large-scale Mixed-Integer Linear Programming (MILP) problem for distributed unit (DU) and centralized unit (CU) placement that minimizes instantaneous power consumption under stringent functional split and service-latency constraints. Similarly, \cite{9392230} proposes an optimization framework for DU/CU placement that jointly considers power consumption and average service-level latency, revealing important trade-offs between energy efficiency and quality of service. Complementing these optimization-based approaches, \cite{10268589} introduces a ML-based algorithm for energy-efficient placement of baseband functions at O-DU/O-CU nodes, illustrating how data-driven methods can further enhance resource utilization and adapt to dynamic traffic conditions in O-RAN deployments.

In \cite{11162430}, throughput-based, end-to-end power consumption models are introduced for O-RANs, capturing how the placement of baseband functions at different O-RAN nodes interacts with multiple functional configuration options \cite{8479363}. Each configuration comes with its own advantages, depending on the mobile network operator’s requirements and deployment strategy (e.g., latency targets, coverage, and cost constraints). Building on that framework, in this work, we go a step further by incorporating application-level processing of ML/AI tasks, together with the associated energy and latency of the inference. This extension allows us to model a complete use-case that jointly accounts for both the O-RAN network behavior and the application-specific ML/AI workload, providing a unified view of end-to-end latency and energy consumption.
In addition, related work such as \cite{crespo2025energyawarecpuorchestrationoran} shows that carefully controlling the CPU frequency can significantly reduce CPU power consumption, an insight we leverage when analyzing the trade-offs between energy efficiency and performance in our extended model.

A number of studies on ML task inference have analysed the energy consumption and latency associated with different models and hardware platforms, while other works have focused on improving network energy efficiency under stringent user-latency requirements by optimally placing baseband functions to meet energy and latency targets. However, these two aspects are typically treated in isolation. Beyond average-latency and energy-aware placement studies, recent works have also explored statistical and closed-loop latency guarantees for edge computing \cite{10129253} and goal-oriented communication systems\cite{10838325}. These studies model communication and computation delays as random variables, showing that designs based only on average latency may under-provision resources for reliability-critical services.  Similarly, work on satellite edge computing\cite{lyholm2026statisticalanalysisenergyefficientsatellite} further shows that frequency-dependent AI inference latency can be modeled statistically to satisfy quantile-based latency guarantees while reducing energy consumption. Although complementary, these works mainly focus on statistical or closed-loop latency characterization. In contrast, our work brings these aspects together by combining a throughput-based, end-to-end energy consumption model for the O-RAN network with an application-oriented view of ML task inference and its latency, allowing us to jointly capture both network-level and ML-level contributions to overall energy use and latency.

We adopt an end-to-end deterministic latency abstraction to evaluate how different O-RAN architectures and AI inference placement decisions affect overall energy consumption and service latency. The formulation is designed at the system level, with communication and computation delays captured through an aggregate deterministic representation rather than through stage-specific component-wise or statistical latency models. Nevertheless, detailed closed-loop or quantile-based latency guarantees are complementary to our model and remain an interesting direction for future extension. In our framework, AI model inference latency is influenced by several factors, including the computational complexity of the model and the underlying hardware performance~\cite{williams2009roofline}. Most critically, it depends on the AI model architecture itself, such as the number of parameters, depth (i.e., number of layers), and overall structural efficiency~\cite{sandler2018mobilenetv2}. Once the number of operations required per inference is known~\cite{9177369}, and the accelerator’s effective compute and energy efficiencies (in ops/s and ops/J, respectively) are characterized, one can directly estimate both the inference latency and the energy consumed per inference under realistic utilization conditions.

\begin{table}[t]
\centering
\caption{\textsc{List of Key Notations}}
\label{tab:notations}
\scriptsize
\setlength{\tabcolsep}{4pt}
\renewcommand{\arraystretch}{1.12}
\begin{tabularx}{\columnwidth}{>{\raggedright\arraybackslash}p{0.27\columnwidth} X}
\toprule
\textbf{Symbol} & \textbf{Description} \\
\midrule

\multicolumn{2}{l}{\textbf{Sets, Indices, and Configuration Definitions}}\\
\midrule
$\mathcal{U}$ & Set of nodal units, $\mathcal{U}=\{\text{O-RU, O-DU, O-CU, DC}\}$. \\
$u$ & Nodal index, $u\in\mathcal{U}$. \\
$\mathcal{I}$ & Set of candidate functional configurations, $\mathcal{I}=\{G,F,M,B\}$. \\
$G,F,M,B$ & Functional configuration with BBP at nodal units $u\in\mathcal{U}$. \\
$x_i$ & Binary configuration selection variable for configuration $i\in\mathcal{I}$. \\
$\chi,\eta,\omega$ & Hierarchical binary configuration indicators, each taking values in $\{0,1\}$. \\
$(\cdot)$ & Generic transport-segment index, $(\cdot)\in\{\mathrm{FH,MH,BH}\}$. \\

\midrule
\multicolumn{2}{l}{\textbf{Latency Terms}}\\
\midrule
$L_i(f,\rho)$ & End-to-end latency under configuration $i$. \\
$L_{\mathrm{o},l}$ & Fiber propagation latency. \\
$L_{\mathrm{sw}}$ & Switching latency. \\
$L_{\mathrm{ro}}$ & Router latency. \\
$L_{\mathrm{en}}$ & \gls{ecpri} encapsulation latency. \\
$L_{\mathrm{bp}}$ & Baseband processing latency. \\
$L_{\mathrm{in}}$ & AI inference latency co-located with BBP. \\

\midrule
\multicolumn{2}{l}{\textbf{Energy Terms}}\\
\midrule
$E_{\mathrm{B},u}$ & BBP energy per bit at node $u$. \\
$E_{\mathrm{I},u}$ & Inference energy per bit at node $u$. \\
$E_{\mathrm{P}}$ & Total processing energy per bit. \\
$E_{\mathrm{Eq}}$ & Node equipment energy per bit. \\
$E_{(\cdot)}$ & Segment-wise transport energy per bit. \\
$E_{\mathrm{R}}$ & Radio energy per bit at O-RU. \\
$E_{\mathrm{Tr}}$ & Transmission energy per bit. \\
$E_{\mathrm{T}}$ & Total transport energy per bit. \\
$E$ & Total energy consumption per bit. \\
$E_i$ & Energy consumed per bit under configuration $i$, where $i\in\mathcal{I}$. \\
$E_{\max}$ & Energy budget constraint. \\

\midrule
\multicolumn{2}{l}{\textbf{Scaling Factors}}\\
\midrule
$\alpha_u,\sigma_u,\gamma_u,\varphi_u$ & Node overprovisioning, overhead, traffic-scaling, and coverage factors. \\
$\alpha_{(\cdot)},\sigma_{(\cdot)},\gamma_{(\cdot)},\phi_{(\cdot)}$ & Segment overprovisioning factor,overhead factor,traffic-scaling,coverage factor. \\

\midrule
\multicolumn{2}{l}{\textbf{User and Equipment Parameters}}\\
\midrule
$N_{sc},P_{sc},C_{sc}$ & Number of cores, power per core, and capacity per core. \\
$P_s,C_s$ & Switch power and capacity. \\
$P_l,C_l$ & IP/WDM line-system power and capacity. \\
$P_r,C_r$ & Router power and capacity. \\
$P_{ls,u},C_{ls,u}$ & Line-system chassis power and capacity at node $u$. \\
$H_{(\cdot),s},H_{(\cdot),l},H_{(\cdot),r}$ & Hop counts of switches, links, and routers per segment. \\
$C_N,C_U$ & \gls{ecpri} traffic rate and user baseband traffic rate. \\
$f$ & CPU frequency. \\
$\rho$ & Network load. \\
\bottomrule
\end{tabularx}
\end{table}

\section{Network Architecture and Performance Metrics }
\begin{figure*}[t] 
    \centering
    \includegraphics[width=7in]{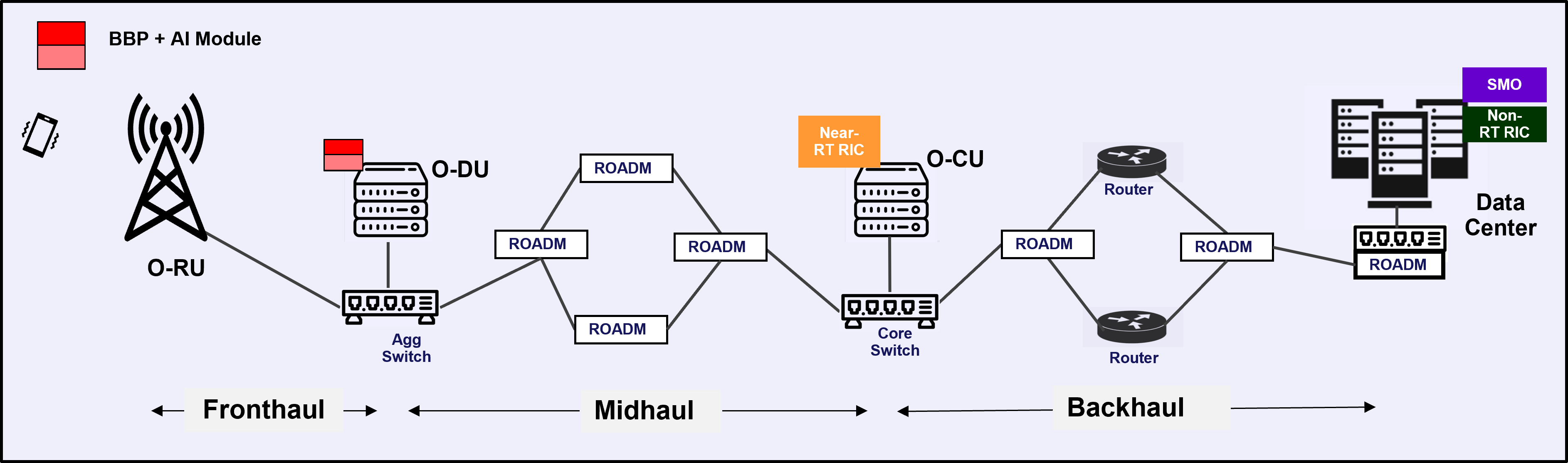}
    \caption{\protect\fontsize{8pt}{9.5pt}\protect\selectfont
    End-to-end O-RAN architecture comprising O-RU, O-DU, O-CU, and DC nodes
    connected through fronthaul (FH), midhaul (MH), and backhaul (BH)
    segments. The total energy and latency are decomposed into processing
    components, namely baseband processing and AI inference, and transport
    components. In the representative configuration shown, both baseband
    processing and AI inference are executed at the O-DU.}
    \label{fig:blockdiagram}
    \vspace{-6mm}
\end{figure*}

In this section, we present the network architecture used in our analysis and also describe the metrics that we have employed for such analysis. The O-RAN architecture in Fig. \ref{fig:blockdiagram} depicts an end-to-end O-RAN view where traffic flows from the \gls{ue} to the O-RU, then through fronthaul (FH) to the O-DU, and onwards via midhaul (MH) and backhaul (BH) to the O-CU and core/data center functions. Along these segments, aggregation/core switches and \gls{roadm}-based transport nodes represent the underlying transport infrastructure, while the Near-RT RIC and the Non-RT RIC/ \gls{smo} in the data center support higher-layer control and orchestration. Consistent with this transport and functional split model \cite{ORANSpec95:online}, the O-RAN architecture comprises three hierarchical network layers: the access network at the user edge (BH), the metro network(MH) that aggregates traffic between radio and centralized units, and the long-haul network/core network(BH) that interconnects metro domains with core and data-center resources

Across this O-RAN processing chain, both baseband processing and application-centric AI/ML tasks can be placed at different nodes. The location where we place the baseband effectively defines the functional split. Pushing it down to the O-RU corresponds to a more distributed split (e.g., Split 2), placing it at the O-DU corresponds to an intermediate split, namely Split 7.2. whereas moving it upward toward the O-CU or even a data center yields a progressively more centralized architecture, approaching high-layer splits such as Split 8. Next, we introduce our energy–latency modeling framework. Specifically, we start from the throughput-based O-RAN energy model in \cite{11162430} and extend it to account for network latency, as well as the energy and inference-latency costs of application-specific AI/ML tasks embedded in the RAN processing pipeline.

\subsection{Energy Consumption} \label{en_model}
In this section, we present the throughput-based energy consumption model for O-RAN. Building on the power consumption model introduced in \cite{11162430}, we incorporate the energy cost of application-specific AI/ML tasks, which are integrated into the RAN processing pipeline. Based on Fig.\ref{fig:blockdiagram}, we structure the end-to-end energy model by decoupling it into two parts: (i) \textit{processing energy}, associated with BBP and AI inference at the nodal units (O-RU/O-DU/O-CU/DC), and (ii) \textit{transport energy}, incurred across the transmission links and equipment.
\vspace{1mm}
\subsubsection{\normalfont\bfseries Processing Energy}
Let  $\mathcal{U}\triangleq$ $\{\textrm{O-RU},\textrm{O-DU},\textrm{O-CU},$ $\textrm{DC}\}$ denote the set of nodes in the O-RAN processing chain. The processing energy at each node $u\in \mathcal{U}$ is comprised of the baseband processing energy ($E_{\mathrm{B},u}$), and the inference energy ($E_{\mathrm{I,u}}$). We define $E_{\mathrm{B},u}$ as the energy consumed by the processing server when baseband processing (BBP) is executed at node $u$. If BBP is not performed at a given node, the signal traverses the node transparently and $E_{\mathrm{B},u}=0$.  The energy  consumed per bit for BBP at unit \(u\in \mathcal{U}\) is given by 
 \begin{equation}
\begin{aligned}
   E_{\mathrm{B},u}=\alpha_u \sigma_u \varphi_u \gamma_u N_{\mathrm{sc}}\frac{P_{\mathrm{sc}}}{C_{\mathrm{sc}}},
\end{aligned}
\label{eq:2}
\end{equation}
where \(\alpha_u\), \(\sigma_u\), and \(\varphi_u\) are the overprovisioning factor, overhead factor, and coverage factor of the unit \(u\), respectively. These factors model the peak-load headroom, site energy overhead, and coverage-driven densification, respectively \cite{11162430}. The factor \(\gamma_u=C_{\mathrm{N}}/C_{\mathrm{U}}\), represents the ratio between the network eCPRI traffic  \(C_{\mathrm{N}}\)
traversing the node and the corresponding user baseband traffic \(C_{\mathrm{U}}\). In addition, \(N_{\mathrm{sc}}\) denotes the number of processing cores at the node, while \(P_{\mathrm{sc}}\) and \(C_{\mathrm{sc}}\) correspond to the power consumption and processing capacity of each core, respectively.

The inference-processing energy ($E_{\mathrm{I,u}}$) is incurred when application-specific AI/ML tasks are executed at the node $u$. The energy consumed per inference is  $E_{\mathrm{in}}=e_{\mathrm{f}}N_{\mathrm{f}}$, where $e_{\mathrm{f}}$ is the hardware energy per floating-point operation and $N_{\mathrm{f}}$ is the number of floating point operations per inference \cite{9177369}. Mapping this cost to the energy consumption model, the inference energy per user bit is given by
 \begin{equation}
\begin{aligned}
  E_{\mathrm{I},u}=\frac{\lambda_{\mathrm{in}} E_{\mathrm{in}}}{C_{\mathrm{U}}},    
\end{aligned}
\label{eq:3}
\end{equation}
where \(\lambda_{\mathrm{in}}\) is the number of inference requests executed per second. Based on edge-AI studies~\cite{8876870} that use AlexNet as a representative image-classification model and evaluate inference throughput, we assume that, on average, one inference request is executed every second by the AI server. Hence, in this work, we use \(\lambda_{\mathrm{in}}=1\). 



Consistent with the baseband energy model, AI/ML inferences are executed only at nodes that perform BBP. So, if BBP is not being performed at node $u$, then it operates in transparent forwarding mode and incurs no inference energy cost such that $E_{\mathrm{I},u}=0$ for such nodes. Consequently, the overall processing energy per bit reflects the combined impact of baseband computation and application-specific inference across the O-RAN architecture. It is expressed as

 \begin{equation}
\begin{aligned}
E_{\mathrm{P}}=\sum_{u\in \mathcal{U}}\left(E_{\mathrm{B},u}+E_{{\mathrm{I}},u}\right).
\end{aligned}
\label{eq:5}
\end{equation}

\subsubsection{\normalfont\bfseries Transport Energy}
The transport energy ($E_{\mathrm{T}}$) consumed in the network is comprised of the equipment energy ($E_{\mathrm{Eq}}$) incurred at the nodal units, and the transmission energy ($E_{\mathrm{Tr}}$) associated with the network links, i.e., $E_{\mathrm{T}}=E_{\mathrm{Eq}}+E_{\mathrm{Tr}}$. The equipment energy accounts for the energy consumed by the transport equipment deployed at node $u$, and depends on the relative placement of the node with respect to the BBP location. This energy component is expressed as
 \begin{equation}
\begin{aligned}
   E_{\mathrm{Eq}}=\sum_{u\in \mathcal{U}}\alpha_u \sigma_u \gamma_u \frac{P_{\mathrm{ls},u}}{C_{\mathrm{ls},u}},    
\end{aligned}
\label{eq:3}
\end{equation}
where \(P_{\mathrm{ls},u}\) and \(C_{\mathrm{ls},u}\) denote the power consumption and capacity of the line-system chassis in the transport equipment at the nodal unit $u\in \mathcal{U}$. When BBP is not executed at node $u$, the traffic scaling factor is
 \(\gamma_u=\varphi_u C_{\mathrm{N}}/C_{\mathrm{U}}\), else, \(\gamma_u=1\).

Transmission energy in the O-RAN system arises from the transport of user traffic across multiple network segments and is influenced by provisioning margins, operational overhead, and traffic aggregation effects. Each segment contributes differently depending on the traversed network elements and their associated power and capacity characteristics. We first consider the backhaul segment, where the cumulative energy contributions of switches, optical links, and routers dominate, and model the corresponding transmission energy per bit as
\begin{equation}
\label{eq:5}
\begin{aligned}
E_{\mathrm{BH}} &=
\alpha_{\mathrm{b}} \sigma_{\mathrm{b}} \gamma_{\mathrm{b}}
\Bigg[
\left(H_{\mathrm{b},\mathrm{s}}+1\right)\frac{P_{\mathrm{s}}}{C_{\mathrm{s}}}
+ H_{\mathrm{b},\mathrm{l}}\frac{P_{\mathrm{l}}}{C_{\mathrm{l}}}
+ \left(H_{\mathrm{b},\mathrm{r}}+1\right)\frac{P_{\mathrm{r}}}{C_{\mathrm{r}}}
\Bigg],
\end{aligned}
\end{equation}
where \(\alpha_{\mathrm{b}}\) and \(\sigma_{\mathrm{b}}\) denote the overprovisioning and overhead factors of the backhaul network, respectively \cite{11162430}.  Moreover, \(H_{\mathrm{b},\mathrm{s}}\), \(H_{\mathrm{b},l}\) and \(H_{\mathrm{b},\mathrm{r}}\) represent the number of hops across switches, \gls{wdm} links, and routers along the backhaul path, while \(P_{\mathrm{s}}\), \(P_{\mathrm{l}}\) and \(P_{\mathrm{r}}\) are the power consumption in the switches, links, and routers, respectively, and \(C_{\mathrm{s}}\), \(C_{\mathrm{l}}\) and \(C_{\mathrm{r}}\) are the corresponding capacities. In addition, \(\gamma_{\mathrm{b}}=\phi_{\mathrm{b}} C_{\mathrm{N}}/C_{\mathrm{U}}\) when BBP is performed at the data center (DC), and \(\gamma_{\mathrm{b}}= 1\) otherwise, where \(\phi_{\mathrm{b}}\) is the coverage factor for backhaul.  Similarly, energy consumption during midhaul transmission is expressed as

\begin{equation}
\label{eq:6}
\begin{aligned}
E_{\mathrm{MH}} &=
\alpha_{\mathrm{m}} \sigma_{\mathrm{m}} \gamma_{\mathrm{m}}
\Bigg[
\left(H_{\mathrm{m},\mathrm{s}}+1\right)\frac{P_{\mathrm{s}}}{C_{\mathrm{s}}}
+ H_{\mathrm{m},\mathrm{l}}\frac{P_{\mathrm{l}}}{C_{\mathrm{l}}}
\Bigg],
\end{aligned}
\end{equation}
where \(\alpha_{\mathrm{m}}\) and \(\sigma_{\mathrm{m}}\) are the overprovisioning factor and overhead factor, respectively, of the midhaul network. Here, \(\gamma_{\mathrm{m}}=\phi_{\mathrm{m}} C_{\mathrm{N}}/C_{\mathrm{U}}\) when BBP is performed at the O-CU or the DC, and \(\gamma_{\mathrm{m}}= 1\) otherwise, with \(\phi_{\mathrm{m}}\) denoting the midhaul coverage factor. Moreover, \(H_{\mathrm{m},\mathrm{s}}\) and \(H_{\mathrm{m},\mathrm{l}}\) are the number of hops across switches and WDM links in the midhaul. Finally, the fronthaul transmission energy per bit is modeled as
\begin{equation}
\label{eq:7}
\begin{aligned}
E_{\mathrm{FH}} &=
\alpha_{\mathrm{f}} \sigma_{\mathrm{f}} \gamma_{\mathrm{f}}
\Bigg[
\left(H_{\mathrm{f},\mathrm{s}}+1\right)\frac{P_{\mathrm{s}}}{C_{\mathrm{s}}} + H_{\mathrm{f},{\mathrm{l}}}\frac{P_{\mathrm{l}}}{C_{\mathrm{l}}}
\Bigg].
\end{aligned}
\end{equation}
Here \(\alpha_{\mathrm{f}}\) and \(\sigma_{\mathrm{f}}\) are the overprovisioning factor and overhead factor, respectively, of the fronthaul network. Moreover, \(\gamma_{\mathrm{f}}= 1\) when BBP is performed at the O-RU, and \(\gamma_{\mathrm{f}}=\phi_{\mathrm{f}} C_{\mathrm{N}}/C_{\mathrm{U}}\) otherwise, with \(\phi_{\mathrm{f}}\) denoting the fronthaul coverage factor. Furthermore,  \(H_{\mathrm{f},\mathrm{s}}\) and \(H_{\mathrm{f},\mathrm{l}}\) denote the number of hops across switches and WDM links in the fronthaul.

In addition to the energy incurred during transmission over DWDM links, the wireless transfer of data from the user device to the radio unit also consumes some energy \(E_{\mathrm{ue}}\), which is the energy expended by the user equipment in transferring one bit of data over the wireless channel to the radio unit. Moreover, the O-RUs incur additional equipment energy given as \(E_{\mathrm{ru}}=N_{\mathrm{ru}} P_{\mathrm{ru}}/C_{\mathrm{U}}\), where \(N_{\mathrm{ru}}\) is the number of O-RUs and \(P_{\mathrm{ru}}\) denotes the static power consumption of the radio hardware at each O-RU. So, the total radio energy consumption is \(E_{\mathrm{R}}=E_{\mathrm{ue}}+E_{\mathrm{ru}}\), which adds to the transmission energy consumption. Therefore, the total transmission energy per bit is given by $E_{\mathrm{Tr}}=E_{\mathrm{R}}+E_{\mathrm{FH}}+E_{\mathrm{MH}}+E_{\mathrm{BH}}$.

Based on the above formulation, the total energy consumption experienced by a user is obtained by combining the energy associated with processing and transport operations. 
Accordingly, the total energy consumption per user is $E=E_{\mathrm{P}}+E_{\mathrm{T}}$.

\subsection{Network Latency } \label{lat_model}
The end-to-end latency of the O-RAN architecture is determined by the combined effects of time consumed during fiber link propagation, traffic switching, eCPRI encapsulation, baseband computation, and AI inference. Accordingly, the overall latency can be decomposed into distinct components, as detailed below.

\subsubsection{\normalfont\bfseries Link Latency}
The time consumed during data transmission depends on the physical distance between the source and destination nodes. The latency during fiber propagation is governed by the propagation delay ($\tau_o$) of the optical fiber. For optical fiber links, the latency of link $l$ is $L_{\mathrm{o},l} = \tau_\mathrm{o} d_l $, where $d_l$ is the length of link $l$. In our work, we assume $\tau_o$ = 5 $\mu$s/km \cite{wang2022edge}. 

\subsubsection{\normalfont\bfseries Traffic Switching Latency} The time required for switching data traffic depends on the switching domain of the traversed element. A ROADM switches traffic at the wavelength level and introduces only a sub-nanosecond delay, which is effectively negligible. On the other hand, an Ethernet switch (E-switch) performs finer, time slot level switching, which includes the optical-electrical-optical (OEO) conversion and electronic forwarding, usually adding a latency of tens of microseconds \cite{wang2022edge}. 

We model a single E-switch with infinite buffer capacity, employing the first-come, first-served (FCFS) principle. Additionally, we model each egress switch port as an M/M/1 queue, with Poisson arrival rate and exponential service rate denoted by $\lambda$ and  $\mu$, respectively \cite{cassandras2008discrete}. The per-port packet service rate $\mu$, measured in packets per second, represents the maximum number of packets that a line card can process per second. For example, in a 10 Gbps switch line card operating with the maximum Ethernet payload size (i.e., jumbo frames), the resulting service rate is $\mu=$ 0.138 Mbps \cite{8567673}.  The mean queuing delay for the M/M/1 system before a switch is given by \cite{cassandras2008discrete}
 \begin{equation}
\begin{aligned}
    L_{\mathrm{q}} = \frac{\rho}{\mu (1- \rho)},    
\end{aligned}  
\end{equation}
where $\rho = \lambda/\mu$ is the normalised load on the switch. The condition $\rho < 1$ ensures that the system always remains stable, whereas larger values of $\rho$ indicate higher traffic conditions leading to higher queuing. As $\rho \rightarrow $ 1, it implies $L_{\mathrm{q}}\rightarrow \infty$ causing packet loss.

The switching latency also includes other components, namely, (i) receiving delay ($ L_{\mathrm{rx}}$), which is the time taken by the switch to buffer the whole packet, (ii) protocol-specific delay ($ L_{\mathrm{ps}}$), which is the time consumed in protocol activities such as header parsing and checks, and (iii) fabric delay ($ L_{\mathrm{sf}}$), which is the time required by the switching fabric to forward the frame. Consequently, the total latency of the E-switch is modeled as $L_{\mathrm{sw}}= L_{\mathrm{q}}+L_{\mathrm{rx}}+L_{\mathrm{ps}}+L_{\mathrm{sf}}$ \cite{han2020evaluation}. When there is no load in the network ($\rho =$ 0), we assume $L_{\mathrm{sw}} =$ 20 $\mu$s \cite{wang2022edge}. We approximate the per-node forwarding delay of intermediate transport elements by a common parameter. So, we assume that the router latency $(L_{\mathrm{ro}})$ and switching latency as comparable, i.e., $L_{\mathrm{ro}} \approx L_{\mathrm{sw}}$. This abstraction is consistent with transport-network delay models that represent intermediate forwarding nodes using uniform per-node delay parameters \cite{ricker2022machine}.

\subsubsection{\normalfont\bfseries Baseband Processing Latency}
The time consumed for BBP at a node mainly depends on the CPU frequency and reduces with the use of a high-frequency CPU in the BBP node. Physical resource blocks (PRBs) and modulation and coding scheme (MCS) affect the computation needed for BBP, hence affecting the latency. As a result, the baseband processing latency is associated with the BBP itself and the hardware properties of the node. It is formulated using a curve-fitted polynomial model as \cite{9392230}

\begin{equation}\label{eq:11}
\begin{aligned}
L_{\mathrm{bp}} =\frac{N_{\mathrm{rb}}}{f^{2}} \sum_{j=0}^{2} \zeta_j{i}^j_{\mathrm{m}},
\end{aligned} 
\end{equation}
where $N_{\mathrm{rb}}$ is the number of PRBs, $f$ is the processing frequency at which the CPU operates, $i_{\mathrm{m}}$ is the MCS index and $\zeta_{j}$ are the polynomial coefficients of the fitted curve. These coefficients have units of $\mu$s when the curve fitting is performed using the values of $L_{\mathrm{bp}}$ and $f$ with units $\mu$s and GHz, respectively \cite{9392230}\cite{8005353}.

\subsubsection{\normalfont\bfseries Interface Encapsulation Latency} When radio data is transmitted on a frame-by-frame basis, the eCPRI encapsulation latency is expressed as $L_{\mathrm{en}} = S_{\mathrm{p}}/R_{\mathrm{l}}$, where $S_{\mathrm{p}}$ is the payload size (in bits) and $R_{\mathrm{l}}$ is the eCPRI line rate option. In our model, we assume $L_{\mathrm{en}} =$  2 $\mu$s\cite{wang2022edge}.

\subsubsection{\normalfont\bfseries AI/ML Model Inference Latency} This refers to the total time taken by an AI/ML model to process a request and generate a response. It depends on the computational complexity, hardware performance, memory access patterns \cite{williams2009roofline}, and most importantly, the AI/ML model's architecture, which includes parameter count, number of layers, and the model structure efficiency \cite{sandler2018mobilenetv2}. The AI/ML inference latency, as characterized in \cite{qi2017paleo}, is decomposed into three components: (i) $L_{\mathrm{rd}}$, which is the time required to fetch (or read) the input parameters, (ii) $L_{\mathrm{ic}}$, which is the time taken to perform the inference computation, and (iii) $L_{\mathrm{wr}}$, which is the time taken to write the output to the local memory. So, the total AI/ML latency is expressed as $L_{\mathrm{in}} = L_{\mathrm{rd}} + L_{\mathrm{ic}} + L_{\mathrm{wr}}$. The time required to transfer input and output data between the device and the memory ($L_{\mathrm{rd}}$ and $L_{\mathrm{wr}}$) is obtained by dividing the total size of the data involved in computation by the available input/output (I/O) bandwidth of the computation device. The computation time ($L_{\mathrm{ic}}$) required to execute the AI/ML inference is determined as $L_{\mathrm{ic}} = N_{\mathrm{op}}/T_{\mathrm{ex}}$, where $N_{\mathrm{op}}$ is the total number of floating-point operations (which quantifies the computational demand of the model) and $T_{\mathrm{ex}}$ is the execution throughput of the device measured in floating-point operations per second \cite{qi2017paleo}. Recent studies analyze LLM inference latency across different serving stages and system-level factors, including request scheduling and batching, prompt prefill, token decoding, and memory-access overhead \cite{10756602}. In this work, these components are not modeled separately; instead, they are abstracted through $L_{\mathrm{rd}}$, $L_{\mathrm{wr}}$, and $L_{\mathrm{ic}}$. For tractability, the analysis focuses on the dominant inference execution cost, i.e., $L_{\mathrm{in}}\approx L_{\mathrm{ic}}$, while detailed token-level serving models are considered complementary.

Using the latency components described above, we obtain the latency in the different parts of the network. The latencies in the fronthaul and midhaul are $L_{\mathrm{FH}}$ and $L_{\mathrm{MH}}$, respectively. These are expressed as $L_{\mathrm{FH}}=\tau_\mathrm{o}d_\mathrm{f}+H_{\mathrm{f},\mathrm{s}}L_{\mathrm{sw}}$ and $L_{\mathrm{MH}}=\tau_\mathrm{o}d_\mathrm{m}+H_{\mathrm{m},\mathrm{s}}L_{\mathrm{sw}}$, where $d_\mathrm{f}$ and $d_\mathrm{m}$ are the total distances in fronthaul and midhaul, respectively. In the backhaul, we encounter added delays due to the presence of routers, so the total latency in the backhaul is $L_{\mathrm{BH}}=\tau_\mathrm{o}d_\mathrm{b}+H_{\mathrm{b},\mathrm{s}}L_{\mathrm{sw}}+H_{\mathrm{b},\mathrm{r}}L_{\mathrm{ro}}$, where $d_\mathrm{b}$ is the total distance in backhaul. Moreover, the latency at a processing nodal unit is $L_\mathrm{N}=L_\mathrm{bp}+L_\mathrm{in}$.

\section{Optimization Problem}
In this section, we explain the optimization problem studied in our work. We first elucidate the proposed configuration-aware models to obtain the energy consumption and latency in O-RAN architectures. We then formulate the optimization problem to determine the location for BBP and AI inference in the network. To run the proposed optimization, the required system parameters are assumed to be known or estimated in advance. These include the candidate configuration set $\mathcal{I}\triangleq\{G,F,M,B\}$, the per-configuration processing and transport energy components ($E_i=E_{\mathrm{P},i}+E_{\mathrm{T},i}$), the per-configuration latency functions $L_i(f,\rho)$, the feasible operating ranges of CPU frequency and network load, i.e., $(f_{\min}\leq f\leq f_{\max})$  and $(0\leq\rho\leq\rho_{\max})$, and the energy budget $E_{\max}$. The AI inference parameters, such as the number of operations per inference and the accelerator compute and energy efficiencies, are also assumed to be characterized beforehand. In this formulation, BBP and AI inference are co-located at the processing node associated with the selected configuration.

\setlength{\textfloatsep}{8pt}
\setlength{\intextsep}{8pt}
\setlength{\abovecaptionskip}{4pt}
\setlength{\belowcaptionskip}{0pt}
\begin{figure}[t!]
    \centering
    \includegraphics[width=3.3in]{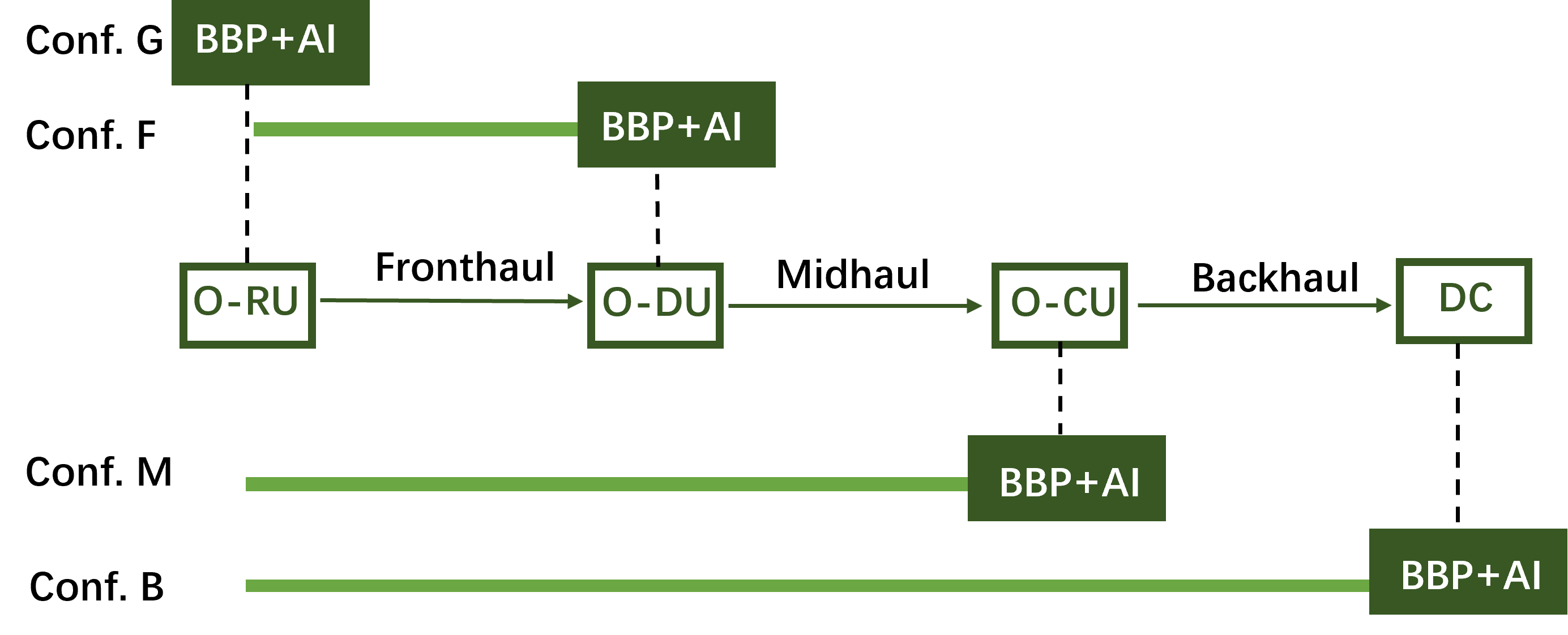}
    \caption{\protect\fontsize{8pt}{9.5pt}\protect\selectfont Representation of the candidate functional configurations $G$, $F$, $M$ and $B$. Each configuration places the BBP and AI inference module at a different O-RAN node, from the O-RU to the DC, resulting in progressively deeper centralization and additional transport segments.} 
    \label{fig:Schematics}
     \vspace{-2mm}
\end{figure}
\vspace{-2mm}

\subsection{Configuration Aware Energy-  Model}
Energy consumption in an O-RAN architecture is influenced by the location at which processing (BBP and AI inference) is executed along the network chain. In Section \ref{en_model}, we model the energy consumption as a function of the node location in an O-RAN network. We now extend this model to an end-to-end energy model under a functional configuration by considering four candidate configurations defined as the set $\mathcal{I}\triangleq\{G,F,M,B\}$, Fig. ~\ref{fig:Schematics} where each element denotes the deepest O-RAN segment included by the selected configuration. Specifically, $G$ corresponds to processing at the O-RU only, $F$ includes the fronthaul up to the processing node O-DU, $M$ includes the fronthaul and midhaul up to the processing node O-CU, and $B$ further includes the backhaul/core with the processing node at DC. In other words, the functional configuration represent the network segment up to the corresponding processing node. Note that processing includes BBP and AI inference. For each candidate configuration $i \in \mathcal{I}$, we model the energy per bit as the sum of processing energy and the transport (transmission) energy. Hence, for each configuration $i$, we can write $E_i=E_{\mathrm{P},i}+E_{\mathrm{T},i}, \; \forall i \in \mathcal{I}$. For example, the energy consumed in configuration $M$, i.e., when BBP is executed at the O-CU is $E_M=E_{\mathrm{P},M}+E_{\mathrm{T},M}$.

The active configuration can be represented by the binary tuple $(\chi,\eta,\omega)\in\{0,1\}^3$, which provides a hierarchical representation of the functional configuration. As such, the total energy per bit is represented as

\begin{equation}\label{eq:energy_eq}
\begin{aligned}
   E(\chi,\eta,\omega) &= \omega \bigl[\eta \bigl(\chi E_G + (1-\chi)E_F \bigr) + (1-\eta)E_M \bigr]\\ &\quad+ (1-\omega)E_B,\\
  & \text{s.t.} \quad  \chi,\eta,\omega \in \{0,1\}.
\end{aligned}
\end{equation}
The binary tuple $(\chi,\eta,\omega)$ is used to encode the configuration choice in a hierarchical way, rather than as three independent decisions. Specifically, $(1,1,1)$ represents configuration $G$, $(0,1,1)$ represents configuration $F$, $(\eta,\omega)=(0,1)$ represents configuration $M$ regardless of the value of $\chi$, and $\omega=0$ represents configuration $B$. In this sense, Eq.~\eqref{eq:energy_eq} provides a compact representation of the four valid configuration options. For the MILP formulation, this logical encoding is then converted into the equivalent one-hot by using Eqs.~\eqref{eq:oh_encoding}-~\eqref{eq:oh_encoding2}.

\subsection{Configuration-Aware Latency Model}
The placement of BBP in an O-RAN architecture is determined by the latency budget associated with the traffic demand. We design an end-to-end latency model under a functional configuration. Accordingly, a latency-aware placement strategy is adopted for BBP to ensure that latency requirements are satisfied. We optimize over the same set of candidate functional configurations $\mathcal{I}=\{G,F,M,B\}$ using a binary tuple $(\chi,\eta,\omega)\in\{0,1\}^3$ that hierarchically activates exactly one configuration in $\mathcal{I}$. 
The resulting end-to-end latency is given by
\begin{equation}\label{eq:13}
\begin{aligned}
   L(\chi,\eta,\omega) &= \omega \bigl[\eta \bigl(\chi L_{\mathrm{G}} + (1-\chi)L_{\mathrm{F}}\bigr) + (1-\eta)L_{\mathrm{M}} \bigr]\\ &\quad + (1-\omega)L_{\mathrm{B}},\\
   &\text{s.t.} \quad  \chi,\eta,\omega \in \{0,1\},
\end{aligned}
\end{equation}
where, $L_{\mathrm{G}} $, $L_{\mathrm{F}}$, $L_{\mathrm{M}}$, and $L_{\mathrm{B}}$ denote the end-to-end latency contributions associated with configurations $G$, $F$, $M$, and $B$, respectively. 
The hierarchical tuple $(\chi,\eta,\omega)$ selects exactly one of $\{L_{\mathrm{G}},L_{\mathrm{F}},L_{\mathrm{M}},L_{\mathrm{B}}\}$, corresponding to progressively deeper functional configurations. The latencies corresponding to each functional configuration are obtained using the latency components described in Section \ref{lat_model}. In configuration $G$, there are delays due to eCPRI encapsulation of user data, and processing at the nodal unit. So, the latency of this configuration is $L_{\mathrm{G}}=L_{\mathrm{en}}+L_{\mathrm{N}}$. The other configurations have additional delays due to transmission over network links. Hence, the latencies in the configurations $F$, $M$ and $B$ are expressed as $L_{\mathrm{F}}=L_{\mathrm{en}}+L_{\mathrm{FH}}+L_{\mathrm{N}}$, $L_{\mathrm{M}}=L_{\mathrm{en}}+L_{\mathrm{FH}}+L_{\mathrm{MH}}+L_{\mathrm{N}}$, and $L_{\mathrm{B}}=L_{\mathrm{en}}+L_{\mathrm{FH}}+L_{\mathrm{MH}}+L_{\mathrm{BH}}+L_{\mathrm{N}}$, respectively.
\subsection{Optimization Framework}
The key design decision is where to execute the required processing along the O-RAN chain (O-RU, O-DU, O-CU, or DC), since this choice determines both the transport delay across fronthaul, midhaul, and backhaul segments and the processing delay at the selected node. Executing processing at deeper nodes can increase latency due to longer transport paths, but typically reduces computing energy because of the higher-capacity processing devices. In contrast, executing processing at edge nodes can increase computing energy due to local execution, while reducing latency by minimizing network transport. To capture this trade-off, we minimize the end-to-end latency subject to a per-delivered-bit energy budget, while allowing the system to operate at different compute and load conditions (e.g., CPU frequency and traffic load). The resulting formulation yields an implementable mixed-integer program (MIP) that selects one functional configuration and its operating point. We jointly select (i) the functional configuration location at which baseband processing and AI inference are executed along the O-RAN chain, and (ii) an operating point that captures the compute and network conditions. The objective is to minimize the end-to-end latency under an energy-per-delivered-bit budget and feasibility constraints. We optimize over the same set of candidate functional configurations $\mathcal{I}=\{G,F,M,B\}$ introduced earlier.

To obtain a linear formulation suitable for MILP optimization, we introduce a one-hot configuration-selection vector
$\mathbf{x} \triangleq [x_G, x_F, x_M, x_B]$ such that $x_i \in \{0,1\}$ and $\sum_{i \in \mathcal{I}} x_i = 1$. Specifically, the tuple $(\chi,\eta,\omega)$ maps to the four configurations as $(1,1,1)\mapsto G$, $(0,1,1)\mapsto F$, $(\eta,\omega)=(0,1)\mapsto M$ (independent of $\chi$), and $\omega=0\mapsto B$, via:
\begin{align} \label{eq:oh_encoding}
& x_{\mathrm{G}} \le \chi,\quad x_{\mathrm{G}} \le \eta,\quad x_{\mathrm{G}} \le \omega,\quad
  x_{\mathrm{G}} \ge \chi + \eta + \omega - 2, \\
& x_{\mathrm{F}} \le 1-\chi,\quad x_{\mathrm{F}} \le \eta,\quad x_{\mathrm{F}} \le \omega,\quad
  x_{\mathrm{F}} \ge \eta + \omega - \chi - 1, \\
& x_{\mathrm{M}} \le 1-\eta,\quad x_{\mathrm{M}} \le \omega,\quad
  x_{\mathrm{M}} \ge \omega - \eta, \\
& x_{\mathrm{B}} \le 1-\omega,\quad
  x_{\mathrm{B}} \ge 1 - \omega, \\
& \sum_{i \in \{G,F,M,B\}} x_i = 1, \\
& x_i \in \{0,1\},\quad \forall\, i \in \{G,F,M,B\}. \label{eq:oh_encoding2}
\end{align}
Under $\mathbf{x}$, the energy selection reduces to $E(\mathbf{x})=\sum_{i\in\mathcal{I}} x_i E_i$. Similarly, the end-to-end latency can be written as
$L(\mathbf{x}) = \sum_{i\in I} x_i L_i$, where $L_i \in \{L_{\mathrm{G}}, L_{\mathrm{F}}, L_{\mathrm{M}}, L_{\mathrm{B}}\}$ denotes the latency associated with configuration $i \in \mathcal{I}$. This formulation enables a unified treatment of latency and energy quantities indexed by the functional configuration $i \in\mathcal{I}$.

Energy enters the formulation as a constraint, while latency depends on the operating variables: the CPU frequency $f$ and the network load $\rho$, constrained by
$f_{\min} \le f \le f_{\max}$ and $0 \le \rho \le \rho_{\max}$. We denote the end-to-end latency associated with configuration $i$ as a function of the operating point $(f,\rho)$. In particular, $L_i(f,\rho)$ captures the BBP latency (as a function of CPU frequency $f$), and the device latency components driven by the load $\rho$ (e.g., switch/router latency). In addition, the propagation term is distance-dominated, eCPRI encapsulation introduces a small line-rate/payload-dependent delay, and AI/ML inference executed on a co-located accelerator can be modeled as a separate fixed compute component. These terms are independent of the baseband CPU frequency and system load. Using the one-hot configuration-selection vector $\mathbf{x}$, the latency minimization under an energy budget is written as
\begin{align}
\min_{\mathbf{x},\, f,\, \rho} \quad
& L(\mathbf{x}, f, \rho)
\;\triangleq\;
\sum_{i \in \mathcal{I}} x_i \, L_i(f,\rho)
\label{eq:opt_obj}
\\
\text{s.t.}\quad
& \sum_{i \in \mathcal{I}} x_i \, E_i \;\le\; E_{\max},
\label{eq:opt_energy}
\\
& \sum_{i \in \mathcal{I}} x_i \;=\; 1,
\label{eq:opt_onehot}
\\
& x_i \in \{0,1\}, \quad \forall i \in \mathcal{I},
\label{eq:opt_binary}
\\
& f_{\min} \le f \le f_{\max},
\label{eq:opt_f}
\\
& 0 \le \rho \le \rho_{\max}.
\label{eq:opt_rho}
\end{align}
Here, $L_i(f,\rho)$ corresponds to the end-to-end latency contribution when configuration $i$ is active, i.e., $L_G(f,\rho)$, $L_F(f,\rho)$, $L_M(f,\rho)$, and $L_B(f,\rho)$.

The formulation \eqref{eq:opt_obj}--\eqref{eq:opt_rho} is an MIP due to the binary configuration-selection variables $x_i \in \{0,1\}$. If $L_i(f,\rho)$ is linear in $(f,\rho)$ for all $i \in\mathcal{I}$, the formulation becomes an MILP. In general, MILPs are NP-hard and may exhibit exponential worst-case complexity as the number of binary variables increases. For any fixed operating point $(f,\rho)$, both the objective
$\sum_{i\in\mathcal{I}} x_i L_i(f,\rho)$ and the energy constraint $\sum_{i\in\mathcal{I}} x_i E_i \le E_{\max}$ are linear in $\mathbf{x}$.
Evidently, the functional configurations set $\mathcal{I}$ has a cardinality $|\mathcal{I}|=$ 4, so the MIP can be solved exactly by enumerating all $i \in\mathcal{I}$. For each configuration $i$, we solve the continuous subproblem in the operating variables $(f,\rho)$ subject to the constraints $f_{\min} \le f \le f_{\max}$ and $0 \le \rho \le \rho_{\max}$, together with the energy constraint $\sum_{i\in\mathcal{I}} x_i E_i \le E_{\max}$, and select the feasible configuration that attains the minimum end-to-end latency. Consequently, the overall runtime scales linearly with $|\mathcal{I}|$ and is dominated by solving at most four low-dimensional continuous optimization problems. If a grid search with $N_f$ frequency samples and $N_\rho$ load samples is used, the computational complexity is $O(|\mathcal{I}|N_fN_\rho)$. Thus, the runtime is dominated by solving at most four low-dimensional continuous minimizations over $(f,\rho)$. This enables efficient re-optimization under changing compute and load conditions.

\section{Results and Discussion}
In this section, we present the results obtained from the proposed energy and latency models and analyze the associated trade-offs. We further examine how users' quality of service (QoS) requirements and network traffic saturation jointly influence the optimal placement of user requests across different baseband processing locations within the O-RAN architecture. Details of the network equipment used in our analysis are provided in \cite{11162430}. We assume that the average monthly data consumption of a user is 40 Gigabytes\cite{Mobileda13:online}. Typical of O-RAN deployments, the servers at O-RU, O-DU, and O-CU have 4 cores with a total BBP capacity of 1 Gbps, whereas the server at DC has 20 cores, which gives a cumulative BBP capacity of 5 Gbps per server \cite{11162430}. The thermal design power per core for the servers at O-RU, O-DU, and O-CU is 6 W, and for those at DC, it is 5.5 W \cite{11162430}. 
Unless stated otherwise, all system parameters are held constant throughout the analysis.

For the AI inference workload, we employ AlexNet \cite{chen2019eyeriss} as a representative convolutional neural network (CNN) model, capturing the computational characteristics of moderate complexity edge vision applications commonly supported in AI-native O-RAN deployments \cite{chen2019eyeriss}. Since inference energy and latency primarily scale with operation count and memory traffic rather than architectural details, AlexNet provides a realistic and reproducible workload for evaluating energy-latency trade-offs without loss of generality. The hardware configurations of the O-RU, O-DU, O-CU, and DC are chosen to reflect representative platforms reported in the literature with parameters adopted from \cite{11162430}. AI/ML inference is assumed to run on a dedicated accelerator co-located with the baseband node and is therefore largely decoupled, and so, we model inference as a fixed (or separately provisioned) compute term. For the O-RU accelerator parameters, we follow \cite{Untitled97:online} while for the O-DU, O-CU, and DC, we draw representative accelerator configurations from \cite{L4Tensor30:online}. For the latency analysis, representative link distances are assumed for the different transport segments of the O-RAN architecture. Specifically, the fronthaul, midhaul, and backhaul links are modeled with lengths of 20 km, 40 km, and 100 km, respectively. These values are selected to reflect typical deployment scenarios and are consistent with prior studies \cite{8479363,9771187}.

\subsection{Energy Latency Tradeoff in Configuration Selection}
We analyse the energy consumption and latency associated with different functional configurations characterized by the location of the processing node (refer Section IV.A). The AI inference accelerators and the BBP units are assumed to be co-located at the same network node, depending on the selected processing location. For simplicity, we consider an unloaded network ($\rho$ = 0) and fix the server frequency at $f$ = 2~GHz. As illustrated in Fig.~\ref{fig:res1}, the energy consumption is significantly higher when processing is performed at the network edge, particularly at the O-RU, compared to processing at the O-DU, O-CU, or the DC. This behavior primarily stems from the use of less energy-efficient processors at the edge and the limited ability to exploit multiplexing gains at the O-RU. In contrast, processing at the O-DU, O-CU, or DC benefits from more centralized computational resources and more efficient processors, leading to lower energy consumption. However, the latency results exhibit an opposite trend. As baseband processing is moved deeper into the network, the cumulative delay introduced by additional network devices and transport links increases, resulting in higher end-to-end latency. Such latency growth is unsuitable for latency-critical services. This analysis highlights the inherent trade-off between energy consumption and latency while choosing the functional configuration in O-RAN deployments.


\input{Figures/Results/fig_energy_latency}

\subsection{End-to-End Latency Analysis} 
We now evaluate the end-to-end latency of the network for different functional configurations. We also study how this latency is affected by variations in load and server frequency. We emulate different operating conditions and obtain the latency in each case. Based on these latency samples $\ell_i$, the empirical cumulative distribution function (CDF) of the latency value $\ell$ is defined as

\begin{equation}\label{cdf}
\mathbb{F}(\ell) = \frac{1}{n} \sum_{i=1}^{n} \mathbb{I} (\ell_i \le \ell)
\end{equation}
where $n$ is the number of latency samples. Moreover, $\mathbb{I}(\cdot)$ is an indicator function such that $\mathbb{I} (\ell_i \le \ell)=$ 1 if $\ell_i \le \ell$, and $\mathbb{I} (\ell_i \le \ell)=$ 0, otherwise. Each point $(\ell, \mathbb{F}(\ell))$ on the CDF curve denotes the fraction of packets $\mathbb{F}(\ell)$ that experience latency below a given threshold $\ell$.
\vspace{3mm}
\subsubsection{\normalfont\bfseries Effect of Network Load}
We vary the network load from $\rho$ = 0 (unloaded case) to $\rho$ = 0.99 (near saturation), and obtain the corresponding values of end-to-end latency. In Fig. \ref{LatVsload_Varfreq_zoom_Combined}, we illustrate the impact of network load on latency for each configuration option at three different values of server frequency, viz., $f$ = 0.8 GHz, $f$ = 2 GHz, and $f$ = 3.6 GHz. For low and moderate load levels, the latency remains almost unchanged, indicating that queuing effects are negligible in this region and that the network operates far from congestion. However, as the load approaches saturation, the latency increases sharply for all frequencies, reflecting the rapid growth of queuing delay near the system capacity limit. In Fig. \ref{fig:res6}, we plot the CDFs of the latency values (using Eqn. \ref{cdf}) when load is varied while operating at three different values of $f$. The trends in this figure corroborate the results depicted in Fig. \ref{LatVsload_Varfreq_zoom_Combined}, and also highlight the effect of varying load using different functional configurations. When configuration $G$ is used, the effect of load is minimal, since the transport network is not involved, and the latency is mainly determined by local processing. Once the transport network is included, the impact of load becomes more visible, and the latency increases accordingly. The latency in configuration $F$ shows only a modest increase, approximately on the order of 1~ms, whereas the latency in configuration $M$ rises more noticeably, by roughly 2 to 2.5~ms. The latency in configuration $B$ exhibits the largest increase and the heaviest tail, with delay values extending well beyond those of the other configurations, particularly at lower server frequencies.
\vspace{2mm}

%
%
\pgfplotsset{compat=1.18}
\usepgfplotslibrary{groupplots}
\begin{figure*}[t!] 
    \centering
    \input{Figures/Results/latency_vs_load_ab_2.tex}
     \vspace{-3mm}
    \caption{\protect\fontsize{8pt}{9.5pt}\protect\selectfont Variation in end-to-end latency with increase in network load $\rho$ at different server frequencies $f$ and functional configurations over (a) the full load range $\rho \in [0.01,0.99]$, and (b) zoomed-in view of the high-load regime $\rho \in [0.80,0.99]$.}
    \label{LatVsload_Varfreq_zoom_Combined}
\end{figure*}



\pgfplotsset{compat=1.18}
\usepgfplotslibrary{groupplots}
\begin{figure*}[t!] 
    \centering
    \hspace*{-0.50cm}   \input{Figures/Results/latency_cdf_three_panels.tex}
     \vspace{-6mm}
    \caption{\protect\fontsize{8pt}{9.5pt}\protect\selectfont Empirical CDF of end-to-end latency under varying network load $\rho$ for each functional configuration at three different server frequencies: (a) $f$ = 0.8~GHz, (b) $f$ = 2~GHz, and (c)$f$ = 3.6~GHz.}
    \vspace{-6mm}
    \label{fig:res6}
\end{figure*}


\usepgfplotslibrary{groupplots}
\begin{figure}[ht!] 
    \centering
    \hspace*{-0.50cm}   \input{Figures/Results/latency_vs_frequency_3loads}
     \vspace{-2mm}
    \caption{\protect\fontsize{8pt}{9.5pt}\protect\selectfont Variation in end-to-end latency with change in server frequencies $f$ at different loading conditions $\rho$ and functional configurations.}
    \label{fig:res7}
\end{figure}
\pgfplotsset{compat=1.18}
\usepgfplotslibrary{groupplots}

\subsubsection{\normalfont\bfseries Effect of Server Frequency}
\vspace{2mm}

We now investigate the effect of changing server frequency on the end-to-end latency of the network. As depicted in subplots (a)-(c) of Fig. \ref{fig:res6}, when the server frequency is increased from 0.8~GHz to 2~GHz and then 3.6~GHz, the curves shift progressively from right to left, denoting a clear reduction in latency. So, increasing the server frequency significantly improves latency performance. To further elucidate this effect, we vary the operating frequency from $f$ = 0.8 GHz to $f$ = 3.6 GHz under three different network load conditions: $\rho=$ 0.01, $\rho=$ 0.97, and $\rho=$ 0.99. Note that, for this analysis, we do not include the mid-load condition around $\rho=$ 0.5, because the latency behavior is broadly similar across non-saturated load conditions and becomes distinctly different only as the network approaches saturation. We plot these results in Fig. \ref{fig:res7} which shows that increasing the baseband server frequency reduces the end-to-end latency across all configuration options. This behavior is expected because increasing the processing frequency shortens the computation time, thereby lowering the overall latency. Moreover, the impact of network load is broadly similar at all frequency settings, which validates the results in Fig. \ref{LatVsload_Varfreq_zoom_Combined}(a). These studies indicate how frequency scaling can compensate for queuing-induced delay under increasing network utilization. For example, from Fig. \ref{fig:res7}, we deduce that, when load increases from $\rho$ = 0.01 to $\rho$ = 0.99, the latency for configuration F can be maintained at $\approx$ 2 ms by increasing the server frequency from $f$ = 1.2 GHz to $f$ = 1.6 GHz.

Fig. \ref{fig:res8} depicts the CDFs for the variations in latency values when server frequency is changed at three different load conditions. Each “step” on the CDF curves corresponds to a different server frequency. As shown in Fig. \ref{fig:res8}(a), at very low load ($\rho=$ 0.01), all four configurations achieve relatively small latencies, but the curves of configuration G and configuration F lie furthest to the left, meaning they can satisfy tight latency requests while in configuration M and configuration B, only a small fraction of operating frequencies can meet the low latency demand. Fig. \ref{fig:res8}(b) corresponds to the near saturation load case ($\rho = $ 0.97). The latency for configuration G remains almost constant under both low-load and near-saturation conditions because processing at the O-RU is largely independent of traffic load. However, there is a significant increase in the latencies for configuration M and configuration B which involve longer transport paths that become more pronounced under near-saturation conditions. At high load ($\rho=$ 0.99), saturation in the network reduces the efficiency of the equipments and causes more queuing delays. The processing at O-RU is not affected by the load, so even at high load, the latency behavior fro configuration G remains unchanged as shown in Fig. \ref{fig:res8}(c), but there is considerable increase in latencies for configuration F and configuration M. Saturation of the network causes significant delays in configuration B when processing is done at DC as it has a combined effect of fronthaul, midhaul and backhaul causing latency to increase dramatically such that even at the best server frequency available of 3.6~GHz, the minimum latency achieved is 7~ms, which is unsuitable for time sensitive applications.
\pgfplotsset{compat=1.18}
\usepgfplotslibrary{groupplots}
\begin{figure*}[t!] 
    \centering
    \hspace*{-0.50cm}   \input{Figures/Results/latency_cdf_over_frequency_three_panels.tex}
    \vspace{-5mm}
    \caption{\protect\fontsize{8pt}{9.5pt}\protect\selectfont Empirical CDF of one-way end-to-end latency across each functional configuration under different server frequencies for three different network load levels: (a) $\rho$ = 0.01, (b) $\rho$ = 0.97, and (c) $\rho$ = 0.99.}
    \label{fig:res8}
    \vspace{-4mm}
\end{figure*}

\subsubsection{\normalfont\bfseries Effect of Functional Configuration}
In Figs. \ref{LatVsload_Varfreq_zoom_Combined} - \ref{fig:res8}, discussed in the previous subsections, the effect of choosing the functional configuration location on latency is clearly visible. When processing remains at the O-RU (configuration G), the latency is the lowest because the user data is handled locally and is, therefore, only marginally affected by the transport network. For BBP at the O-DU (configuration F), only the fronthaul segment is involved. This typically includes a single network switch, and although the delay introduced by one switch is relatively small, it can still add a modest but non-negligible latency for applications with stringent delay requirements. When BBP is moved further toward centralized locations, the transport latency becomes the dominant contributor, as the cumulative delay introduced by switches, routers, and network links begins to outweigh the local processing delay. In configuration M, the midhaul contributes additional delay on top of the fronthaul. In configuration B, the backhaul traverses a larger number of intermediate devices and a more complex transport path, making this option the most sensitive to load and congestion.
\subsubsection{\normalfont\bfseries Feasibility Regions for Latency Budgets}

For a more holistic analysis, we now determine the feasible operating conditions for a given latency budget under different functional configurations in the network. As shown in Fig. \ref{fig:latency_cube}, we examine the latency under different operating conditions characterized by the network load $\rho$ and server frequency $f$. On these latency heatmaps, we illustrate contour lines denoting different representative latency budgets between 0.3 ms and 4 ms. The area above these contours indicates the feasible operating regions that satisfy the corresponding latency constraints. The latency budget contours in Fig. \ref{fig:latency_cube}(a) are all the straight lines as the latency in configuration G (where processing is at the O-RU) does not depend on the network load, but it surely depends on the server baseband frequency. In Fig. \ref{fig:latency_cube}(b), the latency is also affected by the load in the network, and so the budget contours are no longer straight for configuration F. Moreover, the feasibility regions start to shrink because processing at the O-DU incurs higher latency. For configuration M, as shown in Fig. \ref{fig:latency_cube}(c), some of the lower budget contours disappear from the heatmap as they are not achievable even at the maximum server baseband frequency and at the lowest load. This is due to higher processing delays when the BBP is located at the O-CU. This latency further deteriorates as we move towards a more centralized network architecture in configuration B, where the processing is at the DC. As depicted in Fig. \ref{fig:latency_cube}(d), with the highest server frequency and zero load (no queuing delay), the latency is $\sim$1.3~ms. Here, the traffic has to pass through fronthaul,  midhaul, and backhaul of the network, experiencing maximum delays and a long network path. So, we lose most of the feasibility regions when using configuration B, making it unsuitable for latency-constrained workloads with stringent latency demands. Fig. \ref{fig:latency_cube} clearly shows that, for all configurations, we can meet more stringent latency requirements and stricter latency budgets if we increase the server baseband frequency even at higher loads. 

\usepgfplotslibrary{groupplots}
\pgfplotsset{compat=1.18}
\begin{figure*}[t]
    \centering
    \input{Figures/Results/latency_heatmaps_four_panels.tex}
    \caption{\protect\fontsize{8pt}{9.5pt}\protect\selectfont Latency heatmaps for different operating conditions of network load $\rho$ and server frequency $f$ for (a) configuration G, (b) configuration F, (c) configuration M, and (d) configuration B. The heatmaps are overlaid with latency budget contours indicating boundaries of feasibility regions for representative latency requirements.}
    \vspace{-4mm}
    \label{fig:latency_cube}
\end{figure*}

\subsection{Energy-aware Latency Optimization}
We now perform latency optimization which involves minimizing the objective function using Eqn. \ref{eq:opt_obj}. The optimization results in the selection of an optimal configuration corresponding to the lowest achievable latency subject to the constraints given in Eqns. \ref{eq:opt_energy} - \ref{eq:opt_rho}. The choice of the optimal configuration is mainly constrained by the energy budget $E_{\max}$. In Fig.~\ref{engVslat}, we show the results of latency optimization for different representative values of $E_{\max}$ ranging from 7 mJ/bit to 22 mJ/bit. For each value of $E_{\max}$, we denote the optimal configuration. We also plot the energy consumed when operating in the optimal configuration (left axis, in red) and the corresponding optimized latency (right axis, in blue). We infer that a higher energy budget allows the system to move toward more distributed processing (from DC to O-RU), which results in a reduction in latency, but at the expense of higher energy consumption. This highlights the key trade-off in the considered framework, i.e., centralized configurations are more energy efficient, whereas distributed configurations are more suitable for latency-sensitive services.
\input{Figures/Results/optimization}
\section{Conclusion}
In this paper, we present a comprehensive energy and latency analysis of O-RAN architectures with different baseband processing placements and functional configuration options. Using a throughput-based energy consumption model with detailed transmission, processing, and AI inference latency components, we develop a unified framework that captures both network-level and application-level effects. The results demonstrate that placing baseband processing closer to the network edge significantly reduces latency but incurs higher energy consumption due to limited multiplexing gains and lower energy efficiency processors. Conversely, more centralized deployments at the O-CU or data center achieve lower energy per bit at the cost of increased transport latency, which can be prohibitive for time-sensitive applications.
Our evaluation further shows that network load and server operating frequency play a critical role in determining feasibility under strict latency constraints. While increasing baseband processor frequency can compensate for queuing-induced delays under moderate traffic conditions, highly centralized architectures remain vulnerable to congestion effects under heavy load. These findings highlight the necessity of jointly considering energy efficiency, latency budgets, and traffic dynamics when selecting O-RAN functional configurations. Overall, the proposed model provides a systematic tool for guiding baseband and AI inference placement decisions in future O-RAN deployments.
Building on these insights, several directions remain open for further investigation. Future work can further enhance the proposed framework by incorporating resource-sharing factors at different O-RAN nodes, including the O-RU, O-DU, O-CU, and data center, to reflect realistic multi-user and multi-service operation. In addition, examining diverse network node ratios and deployment configurations, such as varying numbers of O-RUs per O-DU or O-DUs per O-CU, would provide deeper insights into aggregation, multiplexing gains, and transport overhead across the network. Another promising direction is to investigate the decoupled placement of baseband processing and application-specific AI workloads, where baseband functions and higher-layer AI tasks may be executed at different network nodes depending on latency tolerance, resource availability, and energy efficiency, while incorporating statistical or quantile-based latency constraints to enable reliability-aware functional configuration and inference placement in O-RAN. Analyzing these placement strategies would help characterize the resulting trade-offs between computation, communication, and inference performance under heterogeneous application demands.
\section*{Acknowledgments}
\small{This work was supported in part by Research Ireland grants 18/CRT/6222 and 13/RC/2077\_P2; EU MSCA grant 101155602; and Horizon Europe SNS “6G-XCEL” Project under Grant 101139194.}

\bibliographystyle{ieeetr}
\bibliography{references}


 




\vfill

\end{document}

%% file: acronyms.tex
\newacronym{3gpp}{3GPP}{3rd Generation Partnership Project}

\newacronym{ai}{AI}{Artificial Intelligence}
\newacronym{aoi}{AoI}{Age of Information}

\newacronym{bs}{BS}{Base Station}
\newacronym{cdf}{CDF}{cumulative distribution function}
\newacronym{cff_frame}{CFC-pull/push}{contention-free pull and contention-push}
\newacronym{cps}{CPS}{cyber-physical systems}
\newacronym{cowu}{CoWu}{Content-based Wake-up}
\newacronym{dl}{DL}{Deep Learning}
\newacronym{dt}{DT}{digital twin}
\newacronym{ecpri}{eCPRI}{enhanced Common Public Radio Interface}
\newacronym{embb}{eMBB}{enhanced Mobile Broadband}
\newacronym{es}{ES}{Edge Server}

\newacronym{gnn}{GNN}{Graph Neural Network}
\newacronym{gNB}{gNB}{next-generation Node B}

\newacronym{harq}{HARQ}{Hybrid Automatic Repeat Request}

\newacronym{iiot}{IIoT}{Industrial Internet of Things}
\newacronym{iot}{IoT}{Internet of Things}

\newacronym{kpi}{KPI}{Key Performance Indicator}
\newacronym{mac}{MAC}{Medium Access Control}
\newacronym{mcu}{MCU}{Micro Controller Unit}
\newacronym{ml}{ML}{Machine Learning}
\newacronym{mmtc}{mMTC}{massive Machine-type Communications}
\newacronym{mse}{MSE}{Meas Square Error}
\newacronym{nrt}{Near-RT}{Near-Real-Time}
\newacronym{near-rt-ric}{Near-RT RIC}{Near-Real-Time RAN Intelligent Controller}
\newacronym{non-rt-ric}{Non-RT RIC}{Non-Real-Time RIC}
\newacronym{o-ran}{O-RAN}{Open-Radio Access Network}
\newacronym{pdcch}{PDCCH}{Physical Downlink Control Channel}
\newacronym{phy}{PHY}{physical}
\newacronym{prach}{PRACH}{Physical Random Access Channel}
\newacronym{pusch}{PUSCH}{Physical Uplink Shared Channel}

\newacronym{qaoi}{QAoI}{Query Age of Information}
\newacronym{qos}{QOS}{quality of servic}
\newacronym{ran}{RAN}{Radio Access Network}
\newacronym{rcs_frame}{RCSC-pull/push}{Reserved pull-contention and shared pull-push contention}
\newacronym{rrc}{RRC}{Radio Resource Control}
\newacronym{ric}{RIC}{RAN Intelligent Controller}
\newacronym{roadm}{ROADM}{reconfigurable optical add-drop multiplexe}
\newacronym{semdas}{SEMDAS}{Semantic Data Sourcing}
\newacronym{snn}{SNNs}{Spiking Neural Networks}
\newacronym{smo}{SMO}{Service Management and Orchestration}
\newacronym{twi}{TWI}{Temporal Windows of Integration}
\newacronym{ucwu}{UCWu}{unicast wake-up}
\newacronym{ue}{UE}{user equipment}
\newacronym{urllc}{URLLC}{Ultra-Reliable and Low Latency Communications}

\newacronym{voi}{VoI}{Value of Information}

\newacronym{wdm}{WDM}{wavelength-division multiplexing}
\newacronym{wsn}{WSN}{Wireless Sensor Network}
\newacronym{wur}{WuR}{Wake-up Radio}
\newacronym{wus}{WuS}{Wake-up Signal}

\newacronym{xr}{XR}{eXtended Reality}


%% file: Figures/Results/fig_energy_latency.tex
\providecommand{\AxisLabelFont}{\fontsize{8}{8}\selectfont}

\begin{figure}[t]
\centering
\begin{tikzpicture}

\begin{axis}[
    name=mainaxis,
    width=0.70\columnwidth,
    height=5cm,
    scale only axis,
    ybar,
    bar width=10pt,
    xmin=0.55, xmax=4.45,
    ymin=0, ymax=14.4,
    xtick={1,2,3,4},
    xticklabels={Conf. G,Conf. F,Conf. M,Conf. B},
    ytick={0,2,4,6,8,10,12,14},
    ylabel={\AxisLabelFont \bfseries Energy (mJ/bit)},
    ylabel style={},
    tick label style={font=\footnotesize},
    xticklabel style={font=\footnotesize},
    axis lines=box,
    major tick length=0pt,
    minor tick length=0pt,
    grid=major,
    major grid style={gray!60,dashed,line width=0.25pt},
    enlarge x limits=0.08,
    bar shift=-5pt,
    point meta=explicit symbolic,
    nodes near coords=\pgfplotspointmeta,
    every node near coord/.append style={
        font=\scriptsize,
        yshift=-3pt
    },
    legend style={
        at={(0.5,0.98)},
        anchor=north,
        draw=none,
        fill=none,
        font=\scriptsize,
        legend columns=2,
        /tikz/every even column/.append style={column sep=0.35cm}
    },
    legend image post style={xscale=1.4,yscale=1.4},
]

\addplot[
    draw=none,
    fill=red
] coordinates {
    (1,13.6) 
    (2,11.8) 
    (3,9.71) 
    (4,6.72) 
};

\end{axis}

\begin{axis}[
    at={(mainaxis.south west)},
    anchor=south west,
    width=0.70\columnwidth,
    height=5cm,
    scale only axis,
    ybar,
    bar width=10pt,
    xmin=0.55, xmax=4.45,
    ymin=0, ymax=1.44,
    xtick=\empty,
    ytick={0,0.2,0.4,0.6,0.8,1.0,1.2,1.4},
    ylabel={\AxisLabelFont \bfseries Latency (ms)},
    ylabel style={},
    tick label style={font=\footnotesize},
    axis y line*=right,
    axis x line=none,
    major tick length=0pt,
    minor tick length=0pt,
    enlarge x limits=0.08,
    bar shift=+5pt,
    point meta=explicit symbolic,
    nodes near coords=\pgfplotspointmeta,
    every node near coord/.append style={
        font=\scriptsize,
        xshift=3pt,
        yshift=-3pt
    }
]

\addplot[
    draw=none,
    fill=blue
] coordinates {
    (1,0.277) 
    (2,0.372) 
    (3,0.632) 
    (4,1.29) 
};

\end{axis}

\node[
    anchor=north,
    yshift=-4pt,
    font=\scriptsize
] at (mainaxis.north) {%
    \begin{tabular}{@{}c@{\hspace{2pt}}l@{\hspace{10pt}}c@{\hspace{2pt}}l@{}}
        \tikz \fill[red] (0,0) rectangle (0.44cm,0.20cm); & Energy &
        \tikz \fill[blue] (0,0) rectangle (0.44cm,0.20cm); & Latency
    \end{tabular}%
};

\end{tikzpicture}
\caption{\protect\fontsize{8pt}{9.5pt}\protect\selectfont Energy and latency for different functional configurations based on the location of the processing node under an unloaded network condition ($\rho = 0$) and fixed server frequency ($f = 2$~GHz).}
\label{fig:res1}
\end{figure}

%% file: Figures/Results/latency_vs_load_ab_2.tex
\begin{tikzpicture}

\definecolor{gray}{RGB}{128,128,128}
\definecolor{green}{RGB}{0,128,0}
\definecolor{purple}{RGB}{128,0,128}

\begin{groupplot}[
    group style={group size=2 by 1},
    clip=true,
    clip mode=individual,
]
\node[font=\scriptsize\bfseries] at (rel axis cs:0.65,-0.20) {(a)};
\nextgroupplot[
height=7cm,
label style={font=\scriptsize\bfseries},
legend cell align={left},
legend cell align={left},
legend columns=2,
legend columns=2,
legend style={
  fill opacity=0.8,
  draw opacity=1,
  text opacity=1,
  at={(0.02,0.98)},
  anchor=north west,
  draw=none
},
legend style={draw=none, fill=none, font=\fontsize{7}{8}\selectfont, at={(0.02,0.98)}, anchor=north west},
major grid style={dashed, gray!60, line width=0.25pt},
tick align=outside,
tick label style={font=\scriptsize\bfseries},
tick pos=left,
width=0.49\textwidth,
x grid style={gray},
xlabel={Load $\rho$},
xmajorgrids,
xmin=-0.03, xmax=1.03,
xtick style={color=black},
y grid style={gray},
ylabel={Latency (ms)},
ylabel style={yshift=-8pt},
ymajorgrids,
ymin=0, ymax=11.5,
ytick style={color=black}
]
\addplot [line width=0.32pt, red, mark=asterisk, mark size=1.75, mark options={solid}]
table {%
0.01 4.18659
0.03 4.18659
0.05 4.18659
0.0699999999999999 4.18659
0.09 4.18659
0.11 4.18659
0.13 4.18659
0.15 4.18659
0.17 4.18659
0.19 4.18659
0.21 4.18659
0.23 4.18659
0.25 4.18659
0.27 4.18659
0.29 4.18659
0.31 4.18659
0.33 4.18659
0.35 4.18659
0.37 4.18659
0.39 4.18659
0.41 4.18659
0.43 4.18659
0.45 4.18659
0.47 4.18659
0.49 4.18659
0.51 4.18659
0.53 4.18659
0.55 4.18659
0.57 4.18659
0.59 4.18659
0.61 4.18659
0.63 4.18659
0.65 4.18659
0.67 4.18659
0.69 4.18659
0.71 4.18659
0.73 4.18659
0.75 4.18659
0.77 4.18659
0.79 4.18659
0.81 4.18659
0.83 4.18659
0.85 4.18659
0.87 4.18659
0.89 4.18659
0.91 4.18659
0.93 4.18659
0.95 4.18659
0.97 4.18659
0.99 4.18659
};
\addlegendentry{Conf. G, $f=$ 0.8 GHz}
\addplot [line width=0.32pt, green, dotted, mark=asterisk, mark size=1.75, mark options={solid}]
table {%
0.01 4.54129278290148
0.03 4.541896458987
0.05 4.54252555301297
0.0699999999999999 4.5431817048465
0.09 4.54386669851887
0.11 4.54458247842371
0.13 4.54533116774946
0.15 4.54611508951406
0.17 4.54693679064082
0.19 4.547799069601
0.21 4.54870500825536
0.23 4.54965800865801
0.25 4.55066183574879
0.27 4.55172066706373
0.29 4.55283915084711
0.31 4.55402247427011
0.33 4.55527644386762
0.35 4.55660758082497
0.37 4.55802323441454
0.39 4.55953171774768
0.41 4.56114247113731
0.43 4.56286625985253
0.45 4.56471541501976
0.47 4.56670412906754
0.49 4.56884882068769
0.51 4.5711685891748
0.53 4.57368578476719
0.55 4.5764267310789
0.57 4.57942264914054
0.59 4.58271085189113
0.61 4.58633630620587
0.63 4.59035370152761
0.65 4.59483022774327
0.67 4.59984936319719
0.69 4.60551612903226
0.71 4.61196451774113
0.73 4.61936822329576
0.75 4.62795652173913
0.77 4.63803843730309
0.79 4.65004071773637
0.81 4.66456979405034
0.83 4.68251747655584
0.85 4.70525120772947
0.87 4.73497993311037
0.89 4.77551910408432
0.91 4.83407568438003
0.93 4.92609316770186
0.95 5.09172463768116
0.97 5.47819806763285
0.99 7.4105652173913
};
\addlegendentry{Conf. M, $f=$ 0.8 GHz}
\addplot [line width=0.32pt, red, mark=o, mark size=1.75, mark options={solid,fill opacity=0}]
table {%
0.01 0.731838
0.03 0.731838
0.05 0.731838
0.0699999999999999 0.731838
0.09 0.731838
0.11 0.731838
0.13 0.731838
0.15 0.731838
0.17 0.731838
0.19 0.731838
0.21 0.731838
0.23 0.731838
0.25 0.731838
0.27 0.731838
0.29 0.731838
0.31 0.731838
0.33 0.731838
0.35 0.731838
0.37 0.731838
0.39 0.731838
0.41 0.731838
0.43 0.731838
0.45 0.731838
0.47 0.731838
0.49 0.731838
0.51 0.731838
0.53 0.731838
0.55 0.731838
0.57 0.731838
0.59 0.731838
0.61 0.731838
0.63 0.731838
0.65 0.731838
0.67 0.731838
0.69 0.731838
0.71 0.731838
0.73 0.731838
0.75 0.731838
0.77 0.731838
0.79 0.731838
0.81 0.731838
0.83 0.731838
0.85 0.731838
0.87 0.731838
0.89 0.731838
0.91 0.731838
0.93 0.731838
0.95 0.731838
0.97 0.731838
0.99 0.731838
};
\addlegendentry{Conf. G, $f=$ 2.0 GHz}
\addplot [line width=0.32pt, green, dotted, mark=o, mark size=1.75, mark options={solid,fill opacity=0}]
table {%
0.01 1.08654078290148
0.03 1.087144458987
0.05 1.08777355301297
0.0699999999999999 1.0884297048465
0.09 1.08911469851887
0.11 1.08983047842371
0.13 1.09057916774946
0.15 1.09136308951407
0.17 1.09218479064082
0.19 1.093047069601
0.21 1.09395300825537
0.23 1.09490600865801
0.25 1.09590983574879
0.27 1.09696866706373
0.29 1.09808715084711
0.31 1.09927047427011
0.33 1.10052444386762
0.35 1.10185558082497
0.37 1.10327123441454
0.39 1.10477971774768
0.41 1.10639047113731
0.43 1.10811425985253
0.45 1.10996341501976
0.47 1.11195212906754
0.49 1.1140968206877
0.51 1.1164165891748
0.53 1.11893378476719
0.55 1.12167473107891
0.57 1.12467064914055
0.59 1.12795885189113
0.61 1.13158430620587
0.63 1.13560170152761
0.65 1.14007822774327
0.67 1.14509736319719
0.69 1.15076412903226
0.71 1.15721251774113
0.73 1.16461622329576
0.75 1.17320452173913
0.77 1.18328643730309
0.79 1.19528871773637
0.81 1.20981779405034
0.83 1.22776547655584
0.85 1.25049920772947
0.87 1.28022793311037
0.89 1.32076710408432
0.91 1.37932368438003
0.93 1.47134116770186
0.95 1.63697263768116
0.97 2.02344606763285
0.99 3.9558132173913
};
\addlegendentry{Conf. M, $f=$ 2.0 GHz}
\addplot [line width=0.32pt, red, mark=square*, mark size=1.75, mark options={solid}]
table {%
0.01 0.276891234567901
0.03 0.276891234567901
0.05 0.276891234567901
0.0699999999999999 0.276891234567901
0.09 0.276891234567901
0.11 0.276891234567901
0.13 0.276891234567901
0.15 0.276891234567901
0.17 0.276891234567901
0.19 0.276891234567901
0.21 0.276891234567901
0.23 0.276891234567901
0.25 0.276891234567901
0.27 0.276891234567901
0.29 0.276891234567901
0.31 0.276891234567901
0.33 0.276891234567901
0.35 0.276891234567901
0.37 0.276891234567901
0.39 0.276891234567901
0.41 0.276891234567901
0.43 0.276891234567901
0.45 0.276891234567901
0.47 0.276891234567901
0.49 0.276891234567901
0.51 0.276891234567901
0.53 0.276891234567901
0.55 0.276891234567901
0.57 0.276891234567901
0.59 0.276891234567901
0.61 0.276891234567901
0.63 0.276891234567901
0.65 0.276891234567901
0.67 0.276891234567901
0.69 0.276891234567901
0.71 0.276891234567901
0.73 0.276891234567901
0.75 0.276891234567901
0.77 0.276891234567901
0.79 0.276891234567901
0.81 0.276891234567901
0.83 0.276891234567901
0.85 0.276891234567901
0.87 0.276891234567901
0.89 0.276891234567901
0.91 0.276891234567901
0.93 0.276891234567901
0.95 0.276891234567901
0.97 0.276891234567901
0.99 0.276891234567901
};
\addlegendentry{Conf. G, $f=$ 3.6 GHz}
\addplot [line width=0.32pt, green, dotted, mark=square*, mark size=1.75, mark options={solid}]
table {%
0.01 0.63159401746938
0.03 0.632197693554903
0.05 0.632826787580868
0.0699999999999999 0.633482939414403
0.09 0.634167933086774
0.11 0.634883712991611
0.13 0.63563240231736
0.15 0.636416324081968
0.17 0.637238025208725
0.19 0.638100304168903
0.21 0.639006242823267
0.23 0.63995924322591
0.25 0.640963070316694
0.27 0.64202190163163
0.29 0.643140385415013
0.31 0.644323708838013
0.33 0.64557767843552
0.35 0.646908815392873
0.37 0.64832446898244
0.39 0.649832952315585
0.41 0.651443705705214
0.43 0.653167494420431
0.45 0.655016649587664
0.47 0.657005363635443
0.49 0.659150055255597
0.51 0.661469823742702
0.53 0.663987019335092
0.55 0.666727965646806
0.57 0.669723883708447
0.59 0.673012086459029
0.61 0.676637540773773
0.63 0.680654936095516
0.65 0.685131462311172
0.67 0.69015059776509
0.69 0.695817363600159
0.71 0.702265752309031
0.73 0.709669457863661
0.75 0.718257756307032
0.77 0.728339671870989
0.79 0.740341952304271
0.81 0.754871028618244
0.83 0.772818711123741
0.85 0.79555244229737
0.87 0.825281167678269
0.89 0.865820338652223
0.91 0.924376918947934
0.93 1.01639440226976
0.95 1.18202587224906
0.97 1.56849930220075
0.99 3.5008664519592
};
\addlegendentry{Conf. M, $f=$ 3.6 GHz}
\addplot [line width=0.32pt, blue, dashed, mark=asterisk, mark size=1.75, mark options={solid}]
table {%
0.01 4.28107319572537
0.03 4.28122411474675
0.05 4.28138138825324
0.0699999999999999 4.28154542621162
0.09 4.28171667462972
0.11 4.28189561960593
0.13 4.28208279193736
0.15 4.28227877237852
0.17 4.28248419766021
0.19 4.28269976740025
0.21 4.28292625206384
0.23 4.2831645021645
0.25 4.2834154589372
0.27 4.28368016676593
0.29 4.28395978771178
0.31 4.28425561856753
0.33 4.2845691109669
0.35 4.28490189520624
0.37 4.28525580860363
0.39 4.28563292943692
0.41 4.28603561778433
0.43 4.28646656496313
0.45 4.28692885375494
0.47 4.28742603226689
0.49 4.28796220517192
0.51 4.2885421472937
0.53 4.2891714461918
0.55 4.28985668276973
0.57 4.29060566228514
0.59 4.29142771297278
0.61 4.29233407655147
0.63 4.2933384253819
0.65 4.29445755693582
0.67 4.2957123407993
0.69 4.29712903225806
0.71 4.29874112943528
0.73 4.30059205582394
0.75 4.30273913043478
0.77 4.30525960932577
0.79 4.30826017943409
0.81 4.31189244851258
0.83 4.31637936913896
0.85 4.32206280193237
0.87 4.32949498327759
0.89 4.33962977602108
0.91 4.35426892109501
0.93 4.37727329192546
0.95 4.41868115942029
0.97 4.51529951690821
0.99 4.99839130434782
};
\addlegendentry{Conf. F, $f=$ 0.8 GHz}
\addplot [line width=0.32pt, purple, dash pattern=on 1pt off 3pt on 3pt off 3pt, mark=asterisk, mark size=1.75, mark options={solid}]
table {%
0.01 5.20158556580296
0.03 5.202792917974
0.05 5.20405110602593
0.0699999999999999 5.205363409693
0.09 5.20673339703774
0.11 5.20816495684742
0.13 5.20966233549892
0.15 5.21123017902813
0.17 5.21287358128165
0.19 5.214598139202
0.21 5.21641001651073
0.23 5.21831601731602
0.25 5.22032367149758
0.27 5.22244133412746
0.29 5.22467830169422
0.31 5.22704494854022
0.33 5.22955288773524
0.35 5.23221516164994
0.37 5.23504646882908
0.39 5.23806343549536
0.41 5.24128494227462
0.43 5.24473251970506
0.45 5.24843083003952
0.47 5.25240825813508
0.49 5.25669764137539
0.51 5.2613371783496
0.53 5.26637156953438
0.55 5.27185346215781
0.57 5.27784529828109
0.59 5.28442170378225
0.61 5.29167261241174
0.63 5.29970740305523
0.65 5.30866045548654
0.67 5.31869872639438
0.69 5.33003225806451
0.71 5.34292903548226
0.73 5.35773644659152
0.75 5.37491304347826
0.77 5.39507687460618
0.79 5.41908143547274
0.81 5.44813958810069
0.83 5.48403495311168
0.85 5.52950241545894
0.87 5.58895986622073
0.89 5.67003820816864
0.91 5.78715136876006
0.93 5.97118633540373
0.95 6.30244927536232
0.97 7.0753961352657
0.99 10.9401304347826
};
\addlegendentry{Conf. B, $f=$ 0.8 GHz}
\addplot [line width=0.32pt, blue, dashed, mark=o, mark size=1.75, mark options={solid,fill opacity=0}]
table {%
0.01 0.82632119572537
0.03 0.826472114746751
0.05 0.826629388253242
0.0699999999999999 0.826793426211626
0.09 0.826964674629718
0.11 0.827143619605928
0.13 0.827330791937365
0.15 0.827526772378517
0.17 0.827732197660206
0.19 0.827947767400251
0.21 0.828174252063842
0.23 0.828412502164502
0.25 0.828663458937198
0.27 0.828928166765932
0.29 0.829207787711778
0.31 0.829503618567528
0.33 0.829817110966905
0.35 0.830149895206243
0.37 0.830503808603635
0.39 0.830880929436921
0.41 0.831283617784328
0.43 0.831714564963133
0.45 0.832176853754941
0.47 0.832674032266886
0.49 0.833210205171924
0.51 0.8337901472937
0.53 0.834419446191798
0.55 0.835104682769726
0.57 0.835853662285137
0.59 0.836675712972782
0.61 0.837582076551468
0.63 0.838586425381904
0.65 0.839705556935818
0.67 0.840960340799297
0.69 0.842377032258065
0.71 0.843989129435283
0.73 0.84584005582394
0.75 0.847987130434783
0.77 0.850507609325772
0.79 0.853508179434093
0.81 0.857140448512586
0.83 0.86162736913896
0.85 0.867310801932367
0.87 0.874742983277592
0.89 0.884877776021081
0.91 0.899516921095008
0.93 0.922521291925466
0.95 0.96392915942029
0.97 1.06054751690821
0.99 1.54363930434783
};
\addlegendentry{Conf. F, $f=$ 2.0 GHz}
\addplot [line width=0.32pt, purple, dash pattern=on 1pt off 3pt on 3pt off 3pt, mark=o, mark size=1.75, mark options={solid,fill opacity=0}]
table {%
0.01 1.74683356580296
0.03 1.748040917974
0.05 1.74929910602593
0.0699999999999999 1.750611409693
0.09 1.75198139703774
0.11 1.75341295684742
0.13 1.75491033549892
0.15 1.75647817902813
0.17 1.75812158128165
0.19 1.759846139202
0.21 1.76165801651073
0.23 1.76356401731602
0.25 1.76557167149758
0.27 1.76768933412746
0.29 1.76992630169422
0.31 1.77229294854022
0.33 1.77480088773524
0.35 1.77746316164994
0.37 1.78029446882908
0.39 1.78331143549537
0.41 1.78653294227463
0.43 1.78998051970506
0.45 1.79367883003953
0.47 1.79765625813508
0.49 1.80194564137539
0.51 1.8065851783496
0.53 1.81161956953438
0.55 1.81710146215781
0.57 1.82309329828109
0.59 1.82966970378226
0.61 1.83692061241174
0.63 1.84495540305523
0.65 1.85390845548654
0.67 1.86394672639438
0.69 1.87528025806452
0.71 1.88817703548226
0.73 1.90298444659152
0.75 1.92016104347826
0.77 1.94032487460618
0.79 1.96432943547274
0.81 1.99338758810069
0.83 2.02928295311168
0.85 2.07475041545894
0.87 2.13420786622074
0.89 2.21528620816864
0.91 2.33239936876007
0.93 2.51643433540373
0.95 2.84769727536232
0.97 3.6206441352657
0.99 7.4853784347826
};
\addlegendentry{Conf. B, $f=$ 2.0 GHz}
\addplot [line width=0.32pt, blue, dashed, mark=square*, mark size=1.75, mark options={solid}]
table {%
0.01 0.371374430293271
0.03 0.371525349314651
0.05 0.371682622821143
0.0699999999999999 0.371846660779526
0.09 0.372017909197619
0.11 0.372196854173828
0.13 0.372384026505266
0.15 0.372580006946418
0.17 0.372785432228107
0.19 0.373001001968152
0.21 0.373227486631743
0.23 0.373465736732403
0.25 0.373716693505099
0.27 0.373981401333833
0.29 0.374261022279679
0.31 0.374556853135429
0.33 0.374870345534806
0.35 0.375203129774144
0.37 0.375557043171536
0.39 0.375934164004822
0.41 0.376336852352229
0.43 0.376767799531034
0.45 0.377230088322842
0.47 0.377727266834787
0.49 0.378263439739825
0.51 0.378843381861601
0.53 0.379472680759699
0.55 0.380157917337627
0.57 0.380906896853038
0.59 0.381728947540683
0.61 0.382635311119369
0.63 0.383639659949805
0.65 0.384758791503719
0.67 0.386013575367198
0.69 0.387430266825966
0.71 0.389042364003183
0.73 0.390893290391841
0.75 0.393040365002684
0.77 0.395560843893673
0.79 0.398561414001994
0.81 0.402193683080487
0.83 0.406680603706861
0.85 0.412364036500268
0.87 0.419796217845493
0.89 0.429931010588981
0.91 0.444570155662909
0.93 0.467574526493367
0.95 0.508982393988191
0.97 0.605600751476114
0.99 1.08869253891573
};
\addlegendentry{Conf. F, $f=$ 3.6 GHz}
\addplot [line width=0.32pt, purple, dash pattern=on 1pt off 3pt on 3pt off 3pt, mark=square*, mark size=1.75, mark options={solid}]
table {%
0.01 1.29188680037086
0.03 1.2930941525419
0.05 1.29435234059384
0.0699999999999999 1.2956646442609
0.09 1.29703463160565
0.11 1.29846619141532
0.13 1.29996357006682
0.15 1.30153141359603
0.17 1.30317481584955
0.19 1.30489937376991
0.21 1.30671125107863
0.23 1.30861725188392
0.25 1.31062490606549
0.27 1.31274256869536
0.29 1.31497953626212
0.31 1.31734618310812
0.33 1.31985412230314
0.35 1.32251639621785
0.37 1.32534770339698
0.39 1.32836467006327
0.41 1.33158617684253
0.43 1.33503375427296
0.45 1.33873206460743
0.47 1.34270949270298
0.49 1.34699887594329
0.51 1.3516384129175
0.53 1.35667280410228
0.55 1.36215469672571
0.57 1.36814653284899
0.59 1.37472293835016
0.61 1.38197384697964
0.63 1.39000863762313
0.65 1.39896169005444
0.67 1.40899996096228
0.69 1.42033349263242
0.71 1.43323027005016
0.73 1.44803768115942
0.75 1.46521427804616
0.77 1.48537810917408
0.79 1.50938267004064
0.81 1.53844082266859
0.83 1.57433618767958
0.85 1.61980365002684
0.87 1.67926110078864
0.89 1.76033944273654
0.91 1.87745260332797
0.93 2.06148756997163
0.95 2.39275050993022
0.97 3.1656973698336
0.99 7.03043166935051
};
\addlegendentry{Conf. B, $f=$ 3.6 GHz}

\nextgroupplot[
height=7cm,
label style={font=\scriptsize\bfseries},
legend cell align={left},
legend cell align={left},
legend columns=2,
legend columns=2,
legend style={
  fill opacity=0.8,
  draw opacity=1,
  text opacity=1,
  at={(0.02,0.98)},
  anchor=north west,
  draw=none
},
legend style={draw=none, fill=none, font=\fontsize{7}{8}\selectfont, at={(0.02,0.98)}, anchor=north west},
major grid style={dashed, gray!60, line width=0.25pt},
tick align=outside,
tick label style={font=\scriptsize\bfseries},
tick pos=left,
width=0.49\textwidth,
x grid style={gray},
xlabel={Load $\rho$},
xmajorgrids,
xmin=0.8, xmax=1,
xtick style={color=black},
y grid style={gray},
ylabel={Latency (ms)},
ylabel style={yshift=-8pt},
ymajorgrids,
ymin=0, ymax=11.5,
ytick style={color=black}
]
\addplot [line width=0.32pt, red, mark=asterisk, mark size=1.75, mark options={solid}]
table {%
0.01 4.18659
0.03 4.18659
0.05 4.18659
0.0699999999999999 4.18659
0.09 4.18659
0.11 4.18659
0.13 4.18659
0.15 4.18659
0.17 4.18659
0.19 4.18659
0.21 4.18659
0.23 4.18659
0.25 4.18659
0.27 4.18659
0.29 4.18659
0.31 4.18659
0.33 4.18659
0.35 4.18659
0.37 4.18659
0.39 4.18659
0.41 4.18659
0.43 4.18659
0.45 4.18659
0.47 4.18659
0.49 4.18659
0.51 4.18659
0.53 4.18659
0.55 4.18659
0.57 4.18659
0.59 4.18659
0.61 4.18659
0.63 4.18659
0.65 4.18659
0.67 4.18659
0.69 4.18659
0.71 4.18659
0.73 4.18659
0.75 4.18659
0.77 4.18659
0.79 4.18659
0.81 4.18659
0.83 4.18659
0.85 4.18659
0.87 4.18659
0.89 4.18659
0.91 4.18659
0.93 4.18659
0.95 4.18659
0.97 4.18659
0.99 4.18659
};
\addlegendentry{Conf. G, $f=$ 0.8 GHz}
\addplot [line width=0.32pt, green, dotted, mark=asterisk, mark size=1.75, mark options={solid}]
table {%
0.01 4.54129278290148
0.03 4.541896458987
0.05 4.54252555301297
0.0699999999999999 4.5431817048465
0.09 4.54386669851887
0.11 4.54458247842371
0.13 4.54533116774946
0.15 4.54611508951406
0.17 4.54693679064082
0.19 4.547799069601
0.21 4.54870500825536
0.23 4.54965800865801
0.25 4.55066183574879
0.27 4.55172066706373
0.29 4.55283915084711
0.31 4.55402247427011
0.33 4.55527644386762
0.35 4.55660758082497
0.37 4.55802323441454
0.39 4.55953171774768
0.41 4.56114247113731
0.43 4.56286625985253
0.45 4.56471541501976
0.47 4.56670412906754
0.49 4.56884882068769
0.51 4.5711685891748
0.53 4.57368578476719
0.55 4.5764267310789
0.57 4.57942264914054
0.59 4.58271085189113
0.61 4.58633630620587
0.63 4.59035370152761
0.65 4.59483022774327
0.67 4.59984936319719
0.69 4.60551612903226
0.71 4.61196451774113
0.73 4.61936822329576
0.75 4.62795652173913
0.77 4.63803843730309
0.79 4.65004071773637
0.81 4.66456979405034
0.83 4.68251747655584
0.85 4.70525120772947
0.87 4.73497993311037
0.89 4.77551910408432
0.91 4.83407568438003
0.93 4.92609316770186
0.95 5.09172463768116
0.97 5.47819806763285
0.99 7.4105652173913
};
\addlegendentry{Conf. M, $f=$ 0.8 GHz}
\addplot [line width=0.32pt, red, mark=o, mark size=1.75, mark options={solid,fill opacity=0}]
table {%
0.01 0.731838
0.03 0.731838
0.05 0.731838
0.0699999999999999 0.731838
0.09 0.731838
0.11 0.731838
0.13 0.731838
0.15 0.731838
0.17 0.731838
0.19 0.731838
0.21 0.731838
0.23 0.731838
0.25 0.731838
0.27 0.731838
0.29 0.731838
0.31 0.731838
0.33 0.731838
0.35 0.731838
0.37 0.731838
0.39 0.731838
0.41 0.731838
0.43 0.731838
0.45 0.731838
0.47 0.731838
0.49 0.731838
0.51 0.731838
0.53 0.731838
0.55 0.731838
0.57 0.731838
0.59 0.731838
0.61 0.731838
0.63 0.731838
0.65 0.731838
0.67 0.731838
0.69 0.731838
0.71 0.731838
0.73 0.731838
0.75 0.731838
0.77 0.731838
0.79 0.731838
0.81 0.731838
0.83 0.731838
0.85 0.731838
0.87 0.731838
0.89 0.731838
0.91 0.731838
0.93 0.731838
0.95 0.731838
0.97 0.731838
0.99 0.731838
};
\addlegendentry{Conf. G, $f=$ 2.0 GHz}
\addplot [line width=0.32pt, green, dotted, mark=o, mark size=1.75, mark options={solid,fill opacity=0}]
table {%
0.01 1.08654078290148
0.03 1.087144458987
0.05 1.08777355301297
0.0699999999999999 1.0884297048465
0.09 1.08911469851887
0.11 1.08983047842371
0.13 1.09057916774946
0.15 1.09136308951407
0.17 1.09218479064082
0.19 1.093047069601
0.21 1.09395300825537
0.23 1.09490600865801
0.25 1.09590983574879
0.27 1.09696866706373
0.29 1.09808715084711
0.31 1.09927047427011
0.33 1.10052444386762
0.35 1.10185558082497
0.37 1.10327123441454
0.39 1.10477971774768
0.41 1.10639047113731
0.43 1.10811425985253
0.45 1.10996341501976
0.47 1.11195212906754
0.49 1.1140968206877
0.51 1.1164165891748
0.53 1.11893378476719
0.55 1.12167473107891
0.57 1.12467064914055
0.59 1.12795885189113
0.61 1.13158430620587
0.63 1.13560170152761
0.65 1.14007822774327
0.67 1.14509736319719
0.69 1.15076412903226
0.71 1.15721251774113
0.73 1.16461622329576
0.75 1.17320452173913
0.77 1.18328643730309
0.79 1.19528871773637
0.81 1.20981779405034
0.83 1.22776547655584
0.85 1.25049920772947
0.87 1.28022793311037
0.89 1.32076710408432
0.91 1.37932368438003
0.93 1.47134116770186
0.95 1.63697263768116
0.97 2.02344606763285
0.99 3.9558132173913
};
\addlegendentry{Conf. M, $f=$ 2.0 GHz}
\addplot [line width=0.32pt, red, mark=square*, mark size=1.75, mark options={solid}]
table {%
0.01 0.276891234567901
0.03 0.276891234567901
0.05 0.276891234567901
0.0699999999999999 0.276891234567901
0.09 0.276891234567901
0.11 0.276891234567901
0.13 0.276891234567901
0.15 0.276891234567901
0.17 0.276891234567901
0.19 0.276891234567901
0.21 0.276891234567901
0.23 0.276891234567901
0.25 0.276891234567901
0.27 0.276891234567901
0.29 0.276891234567901
0.31 0.276891234567901
0.33 0.276891234567901
0.35 0.276891234567901
0.37 0.276891234567901
0.39 0.276891234567901
0.41 0.276891234567901
0.43 0.276891234567901
0.45 0.276891234567901
0.47 0.276891234567901
0.49 0.276891234567901
0.51 0.276891234567901
0.53 0.276891234567901
0.55 0.276891234567901
0.57 0.276891234567901
0.59 0.276891234567901
0.61 0.276891234567901
0.63 0.276891234567901
0.65 0.276891234567901
0.67 0.276891234567901
0.69 0.276891234567901
0.71 0.276891234567901
0.73 0.276891234567901
0.75 0.276891234567901
0.77 0.276891234567901
0.79 0.276891234567901
0.81 0.276891234567901
0.83 0.276891234567901
0.85 0.276891234567901
0.87 0.276891234567901
0.89 0.276891234567901
0.91 0.276891234567901
0.93 0.276891234567901
0.95 0.276891234567901
0.97 0.276891234567901
0.99 0.276891234567901
};
\addlegendentry{Conf. G, $f=$ 3.6 GHz}
\addplot [line width=0.32pt, green, dotted, mark=square*, mark size=1.75, mark options={solid}]
table {%
0.01 0.63159401746938
0.03 0.632197693554903
0.05 0.632826787580868
0.0699999999999999 0.633482939414403
0.09 0.634167933086774
0.11 0.634883712991611
0.13 0.63563240231736
0.15 0.636416324081968
0.17 0.637238025208725
0.19 0.638100304168903
0.21 0.639006242823267
0.23 0.63995924322591
0.25 0.640963070316694
0.27 0.64202190163163
0.29 0.643140385415013
0.31 0.644323708838013
0.33 0.64557767843552
0.35 0.646908815392873
0.37 0.64832446898244
0.39 0.649832952315585
0.41 0.651443705705214
0.43 0.653167494420431
0.45 0.655016649587664
0.47 0.657005363635443
0.49 0.659150055255597
0.51 0.661469823742702
0.53 0.663987019335092
0.55 0.666727965646806
0.57 0.669723883708447
0.59 0.673012086459029
0.61 0.676637540773773
0.63 0.680654936095516
0.65 0.685131462311172
0.67 0.69015059776509
0.69 0.695817363600159
0.71 0.702265752309031
0.73 0.709669457863661
0.75 0.718257756307032
0.77 0.728339671870989
0.79 0.740341952304271
0.81 0.754871028618244
0.83 0.772818711123741
0.85 0.79555244229737
0.87 0.825281167678269
0.89 0.865820338652223
0.91 0.924376918947934
0.93 1.01639440226976
0.95 1.18202587224906
0.97 1.56849930220075
0.99 3.5008664519592
};
\addlegendentry{Conf. M, $f=$ 3.6 GHz}
\addplot [line width=0.32pt, blue, dashed, mark=asterisk, mark size=1.75, mark options={solid}]
table {%
0.01 4.28107319572537
0.03 4.28122411474675
0.05 4.28138138825324
0.0699999999999999 4.28154542621162
0.09 4.28171667462972
0.11 4.28189561960593
0.13 4.28208279193736
0.15 4.28227877237852
0.17 4.28248419766021
0.19 4.28269976740025
0.21 4.28292625206384
0.23 4.2831645021645
0.25 4.2834154589372
0.27 4.28368016676593
0.29 4.28395978771178
0.31 4.28425561856753
0.33 4.2845691109669
0.35 4.28490189520624
0.37 4.28525580860363
0.39 4.28563292943692
0.41 4.28603561778433
0.43 4.28646656496313
0.45 4.28692885375494
0.47 4.28742603226689
0.49 4.28796220517192
0.51 4.2885421472937
0.53 4.2891714461918
0.55 4.28985668276973
0.57 4.29060566228514
0.59 4.29142771297278
0.61 4.29233407655147
0.63 4.2933384253819
0.65 4.29445755693582
0.67 4.2957123407993
0.69 4.29712903225806
0.71 4.29874112943528
0.73 4.30059205582394
0.75 4.30273913043478
0.77 4.30525960932577
0.79 4.30826017943409
0.81 4.31189244851258
0.83 4.31637936913896
0.85 4.32206280193237
0.87 4.32949498327759
0.89 4.33962977602108
0.91 4.35426892109501
0.93 4.37727329192546
0.95 4.41868115942029
0.97 4.51529951690821
0.99 4.99839130434782
};
\addlegendentry{Conf. F, $f=$ 0.8 GHz}
\addplot [line width=0.32pt, purple, dash pattern=on 1pt off 3pt on 3pt off 3pt, mark=asterisk, mark size=1.75, mark options={solid}]
table {%
0.01 5.20158556580296
0.03 5.202792917974
0.05 5.20405110602593
0.0699999999999999 5.205363409693
0.09 5.20673339703774
0.11 5.20816495684742
0.13 5.20966233549892
0.15 5.21123017902813
0.17 5.21287358128165
0.19 5.214598139202
0.21 5.21641001651073
0.23 5.21831601731602
0.25 5.22032367149758
0.27 5.22244133412746
0.29 5.22467830169422
0.31 5.22704494854022
0.33 5.22955288773524
0.35 5.23221516164994
0.37 5.23504646882908
0.39 5.23806343549536
0.41 5.24128494227462
0.43 5.24473251970506
0.45 5.24843083003952
0.47 5.25240825813508
0.49 5.25669764137539
0.51 5.2613371783496
0.53 5.26637156953438
0.55 5.27185346215781
0.57 5.27784529828109
0.59 5.28442170378225
0.61 5.29167261241174
0.63 5.29970740305523
0.65 5.30866045548654
0.67 5.31869872639438
0.69 5.33003225806451
0.71 5.34292903548226
0.73 5.35773644659152
0.75 5.37491304347826
0.77 5.39507687460618
0.79 5.41908143547274
0.81 5.44813958810069
0.83 5.48403495311168
0.85 5.52950241545894
0.87 5.58895986622073
0.89 5.67003820816864
0.91 5.78715136876006
0.93 5.97118633540373
0.95 6.30244927536232
0.97 7.0753961352657
0.99 10.9401304347826
};
\addlegendentry{Conf. B, $f=$ 0.8 GHz}
\addplot [line width=0.32pt, blue, dashed, mark=o, mark size=1.75, mark options={solid,fill opacity=0}]
table {%
0.01 0.82632119572537
0.03 0.826472114746751
0.05 0.826629388253242
0.0699999999999999 0.826793426211626
0.09 0.826964674629718
0.11 0.827143619605928
0.13 0.827330791937365
0.15 0.827526772378517
0.17 0.827732197660206
0.19 0.827947767400251
0.21 0.828174252063842
0.23 0.828412502164502
0.25 0.828663458937198
0.27 0.828928166765932
0.29 0.829207787711778
0.31 0.829503618567528
0.33 0.829817110966905
0.35 0.830149895206243
0.37 0.830503808603635
0.39 0.830880929436921
0.41 0.831283617784328
0.43 0.831714564963133
0.45 0.832176853754941
0.47 0.832674032266886
0.49 0.833210205171924
0.51 0.8337901472937
0.53 0.834419446191798
0.55 0.835104682769726
0.57 0.835853662285137
0.59 0.836675712972782
0.61 0.837582076551468
0.63 0.838586425381904
0.65 0.839705556935818
0.67 0.840960340799297
0.69 0.842377032258065
0.71 0.843989129435283
0.73 0.84584005582394
0.75 0.847987130434783
0.77 0.850507609325772
0.79 0.853508179434093
0.81 0.857140448512586
0.83 0.86162736913896
0.85 0.867310801932367
0.87 0.874742983277592
0.89 0.884877776021081
0.91 0.899516921095008
0.93 0.922521291925466
0.95 0.96392915942029
0.97 1.06054751690821
0.99 1.54363930434783
};
\addlegendentry{Conf. F, $f=$ 2.0 GHz}
\addplot [line width=0.32pt, purple, dash pattern=on 1pt off 3pt on 3pt off 3pt, mark=o, mark size=1.75, mark options={solid,fill opacity=0}]
table {%
0.01 1.74683356580296
0.03 1.748040917974
0.05 1.74929910602593
0.0699999999999999 1.750611409693
0.09 1.75198139703774
0.11 1.75341295684742
0.13 1.75491033549892
0.15 1.75647817902813
0.17 1.75812158128165
0.19 1.759846139202
0.21 1.76165801651073
0.23 1.76356401731602
0.25 1.76557167149758
0.27 1.76768933412746
0.29 1.76992630169422
0.31 1.77229294854022
0.33 1.77480088773524
0.35 1.77746316164994
0.37 1.78029446882908
0.39 1.78331143549537
0.41 1.78653294227463
0.43 1.78998051970506
0.45 1.79367883003953
0.47 1.79765625813508
0.49 1.80194564137539
0.51 1.8065851783496
0.53 1.81161956953438
0.55 1.81710146215781
0.57 1.82309329828109
0.59 1.82966970378226
0.61 1.83692061241174
0.63 1.84495540305523
0.65 1.85390845548654
0.67 1.86394672639438
0.69 1.87528025806452
0.71 1.88817703548226
0.73 1.90298444659152
0.75 1.92016104347826
0.77 1.94032487460618
0.79 1.96432943547274
0.81 1.99338758810069
0.83 2.02928295311168
0.85 2.07475041545894
0.87 2.13420786622074
0.89 2.21528620816864
0.91 2.33239936876007
0.93 2.51643433540373
0.95 2.84769727536232
0.97 3.6206441352657
0.99 7.4853784347826
};
\addlegendentry{Conf. B, $f=$ 2.0 GHz}
\addplot [line width=0.32pt, blue, dashed, mark=square*, mark size=1.75, mark options={solid}]
table {%
0.01 0.371374430293271
0.03 0.371525349314651
0.05 0.371682622821143
0.0699999999999999 0.371846660779526
0.09 0.372017909197619
0.11 0.372196854173828
0.13 0.372384026505266
0.15 0.372580006946418
0.17 0.372785432228107
0.19 0.373001001968152
0.21 0.373227486631743
0.23 0.373465736732403
0.25 0.373716693505099
0.27 0.373981401333833
0.29 0.374261022279679
0.31 0.374556853135429
0.33 0.374870345534806
0.35 0.375203129774144
0.37 0.375557043171536
0.39 0.375934164004822
0.41 0.376336852352229
0.43 0.376767799531034
0.45 0.377230088322842
0.47 0.377727266834787
0.49 0.378263439739825
0.51 0.378843381861601
0.53 0.379472680759699
0.55 0.380157917337627
0.57 0.380906896853038
0.59 0.381728947540683
0.61 0.382635311119369
0.63 0.383639659949805
0.65 0.384758791503719
0.67 0.386013575367198
0.69 0.387430266825966
0.71 0.389042364003183
0.73 0.390893290391841
0.75 0.393040365002684
0.77 0.395560843893673
0.79 0.398561414001994
0.81 0.402193683080487
0.83 0.406680603706861
0.85 0.412364036500268
0.87 0.419796217845493
0.89 0.429931010588981
0.91 0.444570155662909
0.93 0.467574526493367
0.95 0.508982393988191
0.97 0.605600751476114
0.99 1.08869253891573
};
\addlegendentry{Conf. F, $f=$ 3.6 GHz}
\addplot [line width=0.32pt, purple, dash pattern=on 1pt off 3pt on 3pt off 3pt, mark=square*, mark size=1.75, mark options={solid}]
table {%
0.01 1.29188680037086
0.03 1.2930941525419
0.05 1.29435234059384
0.0699999999999999 1.2956646442609
0.09 1.29703463160565
0.11 1.29846619141532
0.13 1.29996357006682
0.15 1.30153141359603
0.17 1.30317481584955
0.19 1.30489937376991
0.21 1.30671125107863
0.23 1.30861725188392
0.25 1.31062490606549
0.27 1.31274256869536
0.29 1.31497953626212
0.31 1.31734618310812
0.33 1.31985412230314
0.35 1.32251639621785
0.37 1.32534770339698
0.39 1.32836467006327
0.41 1.33158617684253
0.43 1.33503375427296
0.45 1.33873206460743
0.47 1.34270949270298
0.49 1.34699887594329
0.51 1.3516384129175
0.53 1.35667280410228
0.55 1.36215469672571
0.57 1.36814653284899
0.59 1.37472293835016
0.61 1.38197384697964
0.63 1.39000863762313
0.65 1.39896169005444
0.67 1.40899996096228
0.69 1.42033349263242
0.71 1.43323027005016
0.73 1.44803768115942
0.75 1.46521427804616
0.77 1.48537810917408
0.79 1.50938267004064
0.81 1.53844082266859
0.83 1.57433618767958
0.85 1.61980365002684
0.87 1.67926110078864
0.89 1.76033944273654
0.91 1.87745260332797
0.93 2.06148756997163
0.95 2.39275050993022
0.97 3.1656973698336
0.99 7.03043166935051
};
\addlegendentry{Conf. B, $f=$ 3.6 GHz}
\node[font=\scriptsize\bfseries] at (rel axis cs:0.5,-0.20) {(b)};
\end{groupplot}

\end{tikzpicture}

%% file: Figures/Results/latency_cdf_three_panels.tex
\begin{tikzpicture}
\definecolor{myblue}{RGB}{31,119,180}
\definecolor{myorange}{RGB}{255,127,14}
\definecolor{mygreen}{RGB}{44,160,44}
\definecolor{myred}{RGB}{214,39,40}
\begin{groupplot}[
    group style={group size=3 by 1, horizontal sep=1cm},
    width=0.37\textwidth,
    height=5.20cm,
    xmin=0, xmax=11,
    xtick={0,2,4,6,8,10},
    ymin=0, ymax=1.05,
    ytick={0,0.2,0.4,0.6,0.8,1.0},
    xlabel={\textbf{Latency (ms)}},
    ylabel={\textbf{CDF}},
      ylabel style={
        at={(axis description cs:-0.10,0.5)},
        anchor=south,
    },
    grid=major,
    major grid style={dashed, gray!55, line width=0.25pt},
    tick label style={font=\scriptsize},
    label style={font=\scriptsize\bfseries},
    legend style={draw=none, fill=none, font=\scriptsize, at={(0.98,0.02)}, anchor=south east},
    legend cell align={left},
    scaled ticks=false,
    clip=false,
]
\nextgroupplot
\addplot+[const plot mark right, color=myblue, mark=asterisk, mark size=1.1pt, line width=0.75pt] coordinates {
(4.18659,0.02)
(4.18659,0.04)
(4.18659,0.06)
(4.18659,0.08)
(4.18659,0.1)
(4.18659,0.12)
(4.18659,0.14)
(4.18659,0.16)
(4.18659,0.18)
(4.18659,0.2)
(4.18659,0.22)
(4.18659,0.24)
(4.18659,0.26)
(4.18659,0.28)
(4.18659,0.3)
(4.18659,0.32)
(4.18659,0.34)
(4.18659,0.36)
(4.18659,0.38)
(4.18659,0.4)
(4.18659,0.42)
(4.18659,0.44)
(4.18659,0.46)
(4.18659,0.48)
(4.18659,0.5)
(4.18659,0.52)
(4.18659,0.54)
(4.18659,0.56)
(4.18659,0.58)
(4.18659,0.6)
(4.18659,0.62)
(4.18659,0.64)
(4.18659,0.66)
(4.18659,0.68)
(4.18659,0.7)
(4.18659,0.72)
(4.18659,0.74)
(4.18659,0.76)
(4.18659,0.78)
(4.18659,0.8)
(4.18659,0.82)
(4.18659,0.84)
(4.18659,0.86)
(4.18659,0.88)
(4.18659,0.9)
(4.18659,0.92)
(4.18659,0.94)
(4.18659,0.96)
(4.18659,0.98)
(4.18659,1)
};
\addlegendentry{Conf. G}
\node[font=\scriptsize\bfseries] at (rel axis cs:0.5,-0.25) {(a)};
\addplot+[const plot mark right, color=myorange, mark=asterisk, mark size=1.1pt, line width=0.75pt] coordinates {
(4.2810732,0.02)
(4.2812241,0.04)
(4.2813814,0.06)
(4.2815454,0.08)
(4.2817167,0.1)
(4.2818956,0.12)
(4.2820828,0.14)
(4.2822788,0.16)
(4.2824842,0.18)
(4.2826998,0.2)
(4.2829263,0.22)
(4.2831645,0.24)
(4.2834155,0.26)
(4.2836802,0.28)
(4.2839598,0.3)
(4.2842556,0.32)
(4.2845691,0.34)
(4.2849019,0.36)
(4.2852558,0.38)
(4.2856329,0.4)
(4.2860356,0.42)
(4.2864666,0.44)
(4.2869289,0.46)
(4.287426,0.48)
(4.2879622,0.5)
(4.2885421,0.52)
(4.2891714,0.54)
(4.2898567,0.56)
(4.2906057,0.58)
(4.2914277,0.6)
(4.2923341,0.62)
(4.2933384,0.64)
(4.2944576,0.66)
(4.2957123,0.68)
(4.297129,0.7)
(4.2987411,0.72)
(4.3005921,0.74)
(4.3027391,0.76)
(4.3052596,0.78)
(4.3082602,0.8)
(4.3118924,0.82)
(4.3163794,0.84)
(4.3220628,0.86)
(4.329495,0.88)
(4.3396298,0.9)
(4.3542689,0.92)
(4.3772733,0.94)
(4.4186812,0.96)
(4.5152995,0.98)
(4.9983913,1)
};
\addlegendentry{Conf. F}
\node[font=\scriptsize\bfseries] at (rel axis cs:0.5,-0.25) {(a)};
\addplot+[const plot mark right, color=mygreen, mark=asterisk, mark size=1.1pt, line width=0.75pt] coordinates {
(4.5412928,0.02)
(4.5418965,0.04)
(4.5425256,0.06)
(4.5431817,0.08)
(4.5438667,0.1)
(4.5445825,0.12)
(4.5453312,0.14)
(4.5461151,0.16)
(4.5469368,0.18)
(4.5477991,0.2)
(4.548705,0.22)
(4.549658,0.24)
(4.5506618,0.26)
(4.5517207,0.28)
(4.5528392,0.3)
(4.5540225,0.32)
(4.5552764,0.34)
(4.5566076,0.36)
(4.5580232,0.38)
(4.5595317,0.4)
(4.5611425,0.42)
(4.5628663,0.44)
(4.5647154,0.46)
(4.5667041,0.48)
(4.5688488,0.5)
(4.5711686,0.52)
(4.5736858,0.54)
(4.5764267,0.56)
(4.5794226,0.58)
(4.5827109,0.6)
(4.5863363,0.62)
(4.5903537,0.64)
(4.5948302,0.66)
(4.5998494,0.68)
(4.6055161,0.7)
(4.6119645,0.72)
(4.6193682,0.74)
(4.6279565,0.76)
(4.6380384,0.78)
(4.6500407,0.8)
(4.6645698,0.82)
(4.6825175,0.84)
(4.7052512,0.86)
(4.7349799,0.88)
(4.7755191,0.9)
(4.8340757,0.92)
(4.9260932,0.94)
(5.0917246,0.96)
(5.4781981,0.98)
(7.4105652,1)
};
\addlegendentry{Conf. M}
\node[font=\scriptsize\bfseries] at (rel axis cs:0.5,-0.25) {(a)};
\addplot+[const plot mark right, color=myred, mark=asterisk, mark size=1.1pt, line width=0.75pt] coordinates {
(5.2015856,0.02)
(5.2027929,0.04)
(5.2040511,0.06)
(5.2053634,0.08)
(5.2067334,0.1)
(5.208165,0.12)
(5.2096623,0.14)
(5.2112302,0.16)
(5.2128736,0.18)
(5.2145981,0.2)
(5.21641,0.22)
(5.218316,0.24)
(5.2203237,0.26)
(5.2224413,0.28)
(5.2246783,0.3)
(5.2270449,0.32)
(5.2295529,0.34)
(5.2322152,0.36)
(5.2350465,0.38)
(5.2380634,0.4)
(5.2412849,0.42)
(5.2447325,0.44)
(5.2484308,0.46)
(5.2524083,0.48)
(5.2566976,0.5)
(5.2613372,0.52)
(5.2663716,0.54)
(5.2718535,0.56)
(5.2778453,0.58)
(5.2844217,0.6)
(5.2916726,0.62)
(5.2997074,0.64)
(5.3086605,0.66)
(5.3186987,0.68)
(5.3300323,0.7)
(5.342929,0.72)
(5.3577364,0.74)
(5.374913,0.76)
(5.3950769,0.78)
(5.4190814,0.8)
(5.4481396,0.82)
(5.484035,0.84)
(5.5295024,0.86)
(5.5889599,0.88)
(5.6700382,0.9)
(5.7871514,0.92)
(5.9711863,0.94)
(6.3024493,0.96)
(7.0753961,0.98)
(10.94013,1)
};
\addlegendentry{Conf. B}
\node[font=\scriptsize\bfseries] at (rel axis cs:0.5,-0.25) {(a)};
\nextgroupplot
\addplot+[const plot mark right, color=myblue, mark=asterisk, mark size=1.1pt, line width=0.75pt] coordinates {
(0.731838,0.02)
(0.731838,0.04)
(0.731838,0.06)
(0.731838,0.08)
(0.731838,0.1)
(0.731838,0.12)
(0.731838,0.14)
(0.731838,0.16)
(0.731838,0.18)
(0.731838,0.2)
(0.731838,0.22)
(0.731838,0.24)
(0.731838,0.26)
(0.731838,0.28)
(0.731838,0.3)
(0.731838,0.32)
(0.731838,0.34)
(0.731838,0.36)
(0.731838,0.38)
(0.731838,0.4)
(0.731838,0.42)
(0.731838,0.44)
(0.731838,0.46)
(0.731838,0.48)
(0.731838,0.5)
(0.731838,0.52)
(0.731838,0.54)
(0.731838,0.56)
(0.731838,0.58)
(0.731838,0.6)
(0.731838,0.62)
(0.731838,0.64)
(0.731838,0.66)
(0.731838,0.68)
(0.731838,0.7)
(0.731838,0.72)
(0.731838,0.74)
(0.731838,0.76)
(0.731838,0.78)
(0.731838,0.8)
(0.731838,0.82)
(0.731838,0.84)
(0.731838,0.86)
(0.731838,0.88)
(0.731838,0.9)
(0.731838,0.92)
(0.731838,0.94)
(0.731838,0.96)
(0.731838,0.98)
(0.731838,1)
};
\addlegendentry{Conf. G}
\node[font=\scriptsize\bfseries] at (rel axis cs:0.5,-0.25) {(b)};
\addplot+[const plot mark right, color=myorange, mark=asterisk, mark size=1.1pt, line width=0.75pt] coordinates {
(0.8263212,0.02)
(0.82647211,0.04)
(0.82662939,0.06)
(0.82679343,0.08)
(0.82696467,0.1)
(0.82714362,0.12)
(0.82733079,0.14)
(0.82752677,0.16)
(0.8277322,0.18)
(0.82794777,0.2)
(0.82817425,0.22)
(0.8284125,0.24)
(0.82866346,0.26)
(0.82892817,0.28)
(0.82920779,0.3)
(0.82950362,0.32)
(0.82981711,0.34)
(0.8301499,0.36)
(0.83050381,0.38)
(0.83088093,0.4)
(0.83128362,0.42)
(0.83171456,0.44)
(0.83217685,0.46)
(0.83267403,0.48)
(0.83321021,0.5)
(0.83379015,0.52)
(0.83441945,0.54)
(0.83510468,0.56)
(0.83585366,0.58)
(0.83667571,0.6)
(0.83758208,0.62)
(0.83858643,0.64)
(0.83970556,0.66)
(0.84096034,0.68)
(0.84237703,0.7)
(0.84398913,0.72)
(0.84584006,0.74)
(0.84798713,0.76)
(0.85050761,0.78)
(0.85350818,0.8)
(0.85714045,0.82)
(0.86162737,0.84)
(0.8673108,0.86)
(0.87474298,0.88)
(0.88487778,0.9)
(0.89951692,0.92)
(0.92252129,0.94)
(0.96392916,0.96)
(1.0605475,0.98)
(1.5436393,1)
};
\addlegendentry{Conf. F}
\node[font=\scriptsize\bfseries] at (rel axis cs:0.5,-0.25) {(b)};
\addplot+[const plot mark right, color=mygreen, mark=asterisk, mark size=1.1pt, line width=0.75pt] coordinates {
(1.0865408,0.02)
(1.0871445,0.04)
(1.0877736,0.06)
(1.0884297,0.08)
(1.0891147,0.1)
(1.0898305,0.12)
(1.0905792,0.14)
(1.0913631,0.16)
(1.0921848,0.18)
(1.0930471,0.2)
(1.093953,0.22)
(1.094906,0.24)
(1.0959098,0.26)
(1.0969687,0.28)
(1.0980872,0.3)
(1.0992705,0.32)
(1.1005244,0.34)
(1.1018556,0.36)
(1.1032712,0.38)
(1.1047797,0.4)
(1.1063905,0.42)
(1.1081143,0.44)
(1.1099634,0.46)
(1.1119521,0.48)
(1.1140968,0.5)
(1.1164166,0.52)
(1.1189338,0.54)
(1.1216747,0.56)
(1.1246706,0.58)
(1.1279589,0.6)
(1.1315843,0.62)
(1.1356017,0.64)
(1.1400782,0.66)
(1.1450974,0.68)
(1.1507641,0.7)
(1.1572125,0.72)
(1.1646162,0.74)
(1.1732045,0.76)
(1.1832864,0.78)
(1.1952887,0.8)
(1.2098178,0.82)
(1.2277655,0.84)
(1.2504992,0.86)
(1.2802279,0.88)
(1.3207671,0.9)
(1.3793237,0.92)
(1.4713412,0.94)
(1.6369726,0.96)
(2.0234461,0.98)
(3.9558132,1)
};
\addlegendentry{Conf. M}
\node[font=\scriptsize\bfseries] at (rel axis cs:0.5,-0.25) {(b)};
\addplot+[const plot mark right, color=myred, mark=asterisk, mark size=1.1pt, line width=0.75pt] coordinates {
(1.7468336,0.02)
(1.7480409,0.04)
(1.7492991,0.06)
(1.7506114,0.08)
(1.7519814,0.1)
(1.753413,0.12)
(1.7549103,0.14)
(1.7564782,0.16)
(1.7581216,0.18)
(1.7598461,0.2)
(1.761658,0.22)
(1.763564,0.24)
(1.7655717,0.26)
(1.7676893,0.28)
(1.7699263,0.3)
(1.7722929,0.32)
(1.7748009,0.34)
(1.7774632,0.36)
(1.7802945,0.38)
(1.7833114,0.4)
(1.7865329,0.42)
(1.7899805,0.44)
(1.7936788,0.46)
(1.7976563,0.48)
(1.8019456,0.5)
(1.8065852,0.52)
(1.8116196,0.54)
(1.8171015,0.56)
(1.8230933,0.58)
(1.8296697,0.6)
(1.8369206,0.62)
(1.8449554,0.64)
(1.8539085,0.66)
(1.8639467,0.68)
(1.8752803,0.7)
(1.888177,0.72)
(1.9029844,0.74)
(1.920161,0.76)
(1.9403249,0.78)
(1.9643294,0.8)
(1.9933876,0.82)
(2.029283,0.84)
(2.0747504,0.86)
(2.1342079,0.88)
(2.2152862,0.9)
(2.3323994,0.92)
(2.5164343,0.94)
(2.8476973,0.96)
(3.6206441,0.98)
(7.4853784,1)
};
\addlegendentry{Conf. B}
\node[font=\scriptsize\bfseries] at (rel axis cs:0.5,-0.25) {(b)};
\nextgroupplot
\addplot+[const plot mark right, color=myblue, mark=asterisk, mark size=1.1pt, line width=0.75pt] coordinates {
(0.27689123,0.02)
(0.27689123,0.04)
(0.27689123,0.06)
(0.27689123,0.08)
(0.27689123,0.1)
(0.27689123,0.12)
(0.27689123,0.14)
(0.27689123,0.16)
(0.27689123,0.18)
(0.27689123,0.2)
(0.27689123,0.22)
(0.27689123,0.24)
(0.27689123,0.26)
(0.27689123,0.28)
(0.27689123,0.3)
(0.27689123,0.32)
(0.27689123,0.34)
(0.27689123,0.36)
(0.27689123,0.38)
(0.27689123,0.4)
(0.27689123,0.42)
(0.27689123,0.44)
(0.27689123,0.46)
(0.27689123,0.48)
(0.27689123,0.5)
(0.27689123,0.52)
(0.27689123,0.54)
(0.27689123,0.56)
(0.27689123,0.58)
(0.27689123,0.6)
(0.27689123,0.62)
(0.27689123,0.64)
(0.27689123,0.66)
(0.27689123,0.68)
(0.27689123,0.7)
(0.27689123,0.72)
(0.27689123,0.74)
(0.27689123,0.76)
(0.27689123,0.78)
(0.27689123,0.8)
(0.27689123,0.82)
(0.27689123,0.84)
(0.27689123,0.86)
(0.27689123,0.88)
(0.27689123,0.9)
(0.27689123,0.92)
(0.27689123,0.94)
(0.27689123,0.96)
(0.27689123,0.98)
(0.27689123,1)
};
\addlegendentry{Conf. G}
\node[font=\scriptsize\bfseries] at (rel axis cs:0.5,-0.25) {(c)};
\addplot+[const plot mark right, color=myorange, mark=asterisk, mark size=1.1pt, line width=0.75pt] coordinates {
(0.37137443,0.02)
(0.37152535,0.04)
(0.37168262,0.06)
(0.37184666,0.08)
(0.37201791,0.1)
(0.37219685,0.12)
(0.37238403,0.14)
(0.37258001,0.16)
(0.37278543,0.18)
(0.373001,0.2)
(0.37322749,0.22)
(0.37346574,0.24)
(0.37371669,0.26)
(0.3739814,0.28)
(0.37426102,0.3)
(0.37455685,0.32)
(0.37487035,0.34)
(0.37520313,0.36)
(0.37555704,0.38)
(0.37593416,0.4)
(0.37633685,0.42)
(0.3767678,0.44)
(0.37723009,0.46)
(0.37772727,0.48)
(0.37826344,0.5)
(0.37884338,0.52)
(0.37947268,0.54)
(0.38015792,0.56)
(0.3809069,0.58)
(0.38172895,0.6)
(0.38263531,0.62)
(0.38363966,0.64)
(0.38475879,0.66)
(0.38601358,0.68)
(0.38743027,0.7)
(0.38904236,0.72)
(0.39089329,0.74)
(0.39304037,0.76)
(0.39556084,0.78)
(0.39856141,0.8)
(0.40219368,0.82)
(0.4066806,0.84)
(0.41236404,0.86)
(0.41979622,0.88)
(0.42993101,0.9)
(0.44457016,0.92)
(0.46757453,0.94)
(0.50898239,0.96)
(0.60560075,0.98)
(1.0886925,1)
};
\addlegendentry{Conf. F}
\node[font=\scriptsize\bfseries] at (rel axis cs:0.5,-0.25) {(c)};
\addplot+[const plot mark right, color=mygreen, mark=asterisk, mark size=1.1pt, line width=0.75pt] coordinates {
(0.63159402,0.02)
(0.63219769,0.04)
(0.63282679,0.06)
(0.63348294,0.08)
(0.63416793,0.1)
(0.63488371,0.12)
(0.6356324,0.14)
(0.63641632,0.16)
(0.63723803,0.18)
(0.6381003,0.2)
(0.63900624,0.22)
(0.63995924,0.24)
(0.64096307,0.26)
(0.6420219,0.28)
(0.64314039,0.3)
(0.64432371,0.32)
(0.64557768,0.34)
(0.64690882,0.36)
(0.64832447,0.38)
(0.64983295,0.4)
(0.65144371,0.42)
(0.65316749,0.44)
(0.65501665,0.46)
(0.65700536,0.48)
(0.65915006,0.5)
(0.66146982,0.52)
(0.66398702,0.54)
(0.66672797,0.56)
(0.66972388,0.58)
(0.67301209,0.6)
(0.67663754,0.62)
(0.68065494,0.64)
(0.68513146,0.66)
(0.6901506,0.68)
(0.69581736,0.7)
(0.70226575,0.72)
(0.70966946,0.74)
(0.71825776,0.76)
(0.72833967,0.78)
(0.74034195,0.8)
(0.75487103,0.82)
(0.77281871,0.84)
(0.79555244,0.86)
(0.82528117,0.88)
(0.86582034,0.9)
(0.92437692,0.92)
(1.0163944,0.94)
(1.1820259,0.96)
(1.5684993,0.98)
(3.5008665,1)
};
\addlegendentry{Conf. M}
\node[font=\scriptsize\bfseries] at (rel axis cs:0.5,-0.25) {(c)};
\addplot+[const plot mark right, color=myred, mark=asterisk, mark size=1.1pt, line width=0.75pt] coordinates {
(1.2918868,0.02)
(1.2930942,0.04)
(1.2943523,0.06)
(1.2956646,0.08)
(1.2970346,0.1)
(1.2984662,0.12)
(1.2999636,0.14)
(1.3015314,0.16)
(1.3031748,0.18)
(1.3048994,0.2)
(1.3067113,0.22)
(1.3086173,0.24)
(1.3106249,0.26)
(1.3127426,0.28)
(1.3149795,0.3)
(1.3173462,0.32)
(1.3198541,0.34)
(1.3225164,0.36)
(1.3253477,0.38)
(1.3283647,0.4)
(1.3315862,0.42)
(1.3350338,0.44)
(1.3387321,0.46)
(1.3427095,0.48)
(1.3469989,0.5)
(1.3516384,0.52)
(1.3566728,0.54)
(1.3621547,0.56)
(1.3681465,0.58)
(1.3747229,0.6)
(1.3819738,0.62)
(1.3900086,0.64)
(1.3989617,0.66)
(1.409,0.68)
(1.4203335,0.7)
(1.4332303,0.72)
(1.4480377,0.74)
(1.4652143,0.76)
(1.4853781,0.78)
(1.5093827,0.8)
(1.5384408,0.82)
(1.5743362,0.84)
(1.6198037,0.86)
(1.6792611,0.88)
(1.7603394,0.9)
(1.8774526,0.92)
(2.0614876,0.94)
(2.3927505,0.96)
(3.1656974,0.98)
(7.0304317,1)
};
\addlegendentry{Conf. B}
\node[font=\scriptsize\bfseries] at (rel axis cs:0.5,-0.25) {(c)};
\end{groupplot}
\end{tikzpicture}

%% file: Figures/Results/latency_vs_frequency_3loads.tex
\begin{tikzpicture}

\definecolor{gray}{RGB}{128,128,128}
\definecolor{green}{RGB}{0,128,0}
\definecolor{purple}{RGB}{128,0,128}

\begin{axis}[
height=7cm,
label style={font=\scriptsize\bfseries},
legend cell align={left},
legend cell align={left},
legend columns=2,
legend columns=2,
legend style={draw=none, fill=white, font=\fontsize{7}{7}\selectfont, at={(1,1)}, anchor=north east},
legend style={fill opacity=1, draw opacity=1, text opacity=1, at={(1,1)}},
major grid style={dashed, gray!60, line width=0.25pt},
tick align=outside,
tick label style={font=\scriptsize},
tick pos=left,
width=1\columnwidth,
x grid style={gray},
xlabel={Server frequency, \(\displaystyle f\) (GHz)},
xmajorgrids,
xmin=0.72, xmax=3.74,
xtick style={color=black},
y grid style={gray},
ylabel={Latency (ms)},
ymajorgrids,
ymin=0, ymax=14,
ytick={0,2,4,6,8,10,12,14},
ytick style={color=black}
]
\addplot [line width=0.32pt, red, mark=asterisk, mark size=1.75, mark options={solid}]
table {%
0.8 4.18659
1 2.705982
1.2 1.90170111111111
1.6 1.10199
1.8 0.886194938271605
2 0.731838
2.2 0.61763132231405
2.6 0.463167514792899
2.8 0.409528775510204
3 0.366255777777778
3.2 0.33084
3.4 0.301488269896194
3.6 0.276891234567901
};
\addlegendentry{Conf. G, $\rho=0.01$}
\addplot [line width=0.32pt, blue, dashed, mark=asterisk, mark size=1.75, mark options={solid}]
table {%
0.8 4.28107319572537
1 2.80046519572537
1.2 1.99618430683648
1.6 1.19647319572537
1.8 0.980678133996975
2 0.82632119572537
2.2 0.712114518039419
2.6 0.557650710518269
2.8 0.504011971235574
3 0.460738973503147
3.2 0.425323195725369
3.4 0.395971465621563
3.6 0.371374430293271
};
\addlegendentry{Conf. F, $\rho=0.01$}
\addplot [line width=0.32pt, red, mark=o, mark size=1.75, mark options={solid,fill opacity=0}]
table {%
0.8 4.18659
1 2.705982
1.2 1.90170111111111
1.6 1.10199
1.8 0.886194938271605
2 0.731838
2.2 0.61763132231405
2.6 0.463167514792899
2.8 0.409528775510204
3 0.366255777777778
3.2 0.33084
3.4 0.301488269896194
3.6 0.276891234567901
};
\addlegendentry{Conf. G, $\rho=0.97$}
\addplot [line width=0.32pt, blue, dashed, mark=o, mark size=1.75, mark options={solid,fill opacity=0}]
table {%
0.8 4.51529951690821
1 3.03469151690821
1.2 2.23041062801932
1.6 1.43069951690821
1.8 1.21490445517982
2 1.06054751690821
2.2 0.946340839222262
2.6 0.791877031701112
2.8 0.738238292418417
3 0.69496529468599
3.2 0.659549516908212
3.4 0.630197786804406
3.6 0.605600751476114
};
\addlegendentry{Conf. F, $\rho=0.97$}
\addplot [line width=0.32pt, red, mark=square*, mark size=1.75, mark options={solid}]
table {%
0.8 4.18659
1 2.705982
1.2 1.90170111111111
1.6 1.10199
1.8 0.886194938271605
2 0.731838
2.2 0.61763132231405
2.6 0.463167514792899
2.8 0.409528775510204
3 0.366255777777778
3.2 0.33084
3.4 0.301488269896194
3.6 0.276891234567901
};
\addlegendentry{Conf. G, $\rho=0.99$}
\addplot [line width=0.32pt, blue, dashed, mark=square*, mark size=1.75, mark options={solid}]
table {%
0.8 4.99839130434782
1 3.51778330434783
1.2 2.71350241545894
1.6 1.91379130434783
1.8 1.69799624261943
2 1.54363930434783
2.2 1.42943262666188
2.6 1.27496881914072
2.8 1.22133007985803
3 1.1780570821256
3.2 1.14264130434783
3.4 1.11328957424402
3.6 1.08869253891573
};
\addlegendentry{Conf. F, $\rho=0.99$}
\addplot [line width=0.32pt, green, dotted, mark=asterisk, mark size=1.75, mark options={solid}]
table {%
0.8 4.54129278290148
1 3.06068478290148
1.2 2.25640389401259
1.6 1.45669278290148
1.8 1.24089772117308
2 1.08654078290148
2.2 0.972334105215528
2.6 0.817870297694378
2.8 0.764231558411683
3 0.720958560679256
3.2 0.685542782901479
3.4 0.656191052797672
3.6 0.63159401746938
};
\addlegendentry{Conf. M, $\rho=0.01$}
\addplot [line width=0.32pt, purple, dash pattern=on 1pt off 3pt on 3pt off 3pt, mark=asterisk, mark size=1.75, mark options={solid}]
table {%
0.8 5.20158556580296
1 3.72097756580296
1.2 2.91669667691407
1.6 2.11698556580296
1.8 1.90119050407456
2 1.74683356580296
2.2 1.63262688811701
2.6 1.47816308059586
2.8 1.42452434131316
3 1.38125134358073
3.2 1.34583556580296
3.4 1.31648383569915
3.6 1.29188680037086
};
\addlegendentry{Conf. B, $\rho=0.01$}
\addplot [line width=0.32pt, green, dotted, mark=o, mark size=1.75, mark options={solid,fill opacity=0}]
table {%
0.8 5.47819806763285
1 3.99759006763285
1.2 3.19330917874396
1.6 2.39359806763285
1.8 2.17780300590445
2 2.02344606763285
2.2 1.9092393899469
2.6 1.75477558242575
2.8 1.70113684314305
3 1.65786384541063
3.2 1.62244806763285
3.4 1.59309633752904
3.6 1.56849930220075
};
\addlegendentry{Conf. M, $\rho=0.97$}
\addplot [line width=0.32pt, purple, dash pattern=on 1pt off 3pt on 3pt off 3pt, mark=o, mark size=1.75, mark options={solid,fill opacity=0}]
table {%
0.8 7.0753961352657
1 5.5947881352657
1.2 4.79050724637681
1.6 3.9907961352657
1.8 3.7750010735373
2 3.6206441352657
2.2 3.50643745757975
2.6 3.3519736500586
2.8 3.2983349107759
3 3.25506191304348
3.2 3.2196461352657
3.4 3.19029440516189
3.6 3.1656973698336
};
\addlegendentry{Conf. B, $\rho=0.97$}
\addplot [line width=0.32pt, green, dotted, mark=square*, mark size=1.75, mark options={solid}]
table {%
0.8 7.4105652173913
1 5.9299572173913
1.2 5.12567632850241
1.6 4.3259652173913
1.8 4.11017015566291
2 3.9558132173913
2.2 3.84160653970535
2.6 3.6871427321842
2.8 3.63350399290151
3 3.59023099516908
3.2 3.5548152173913
3.4 3.5254634872875
3.6 3.5008664519592
};
\addlegendentry{Conf. M, $\rho=0.99$}
\addplot [line width=0.32pt, purple, dash pattern=on 1pt off 3pt on 3pt off 3pt, mark=square*, mark size=1.75, mark options={solid}]
table {%
0.8 10.9401304347826
1 9.45952243478261
1.2 8.65524154589372
1.6 7.85553043478261
1.8 7.63973537305421
2 7.4853784347826
2.2 7.37117175709665
2.6 7.2167079495755
2.8 7.16306921029281
3 7.11979621256038
3.2 7.0843804347826
3.4 7.0550287046788
3.6 7.03043166935051
};
\addlegendentry{Conf. B, $\rho=0.99$}
\end{axis}

\end{tikzpicture}

%% file: Figures/Results/latency_cdf_over_frequency_three_panels.tex
\begin{tikzpicture}
\definecolor{myblue}{RGB}{31,119,180}
\definecolor{myorange}{RGB}{255,127,14}
\definecolor{mygreen}{RGB}{44,160,44}
\definecolor{myred}{RGB}{214,39,40}
\begin{groupplot}[
    group style={group size=3 by 1, horizontal sep=1cm},
    width=0.37\textwidth,
    height=5.20cm,
    xmin=0, xmax=12,
    xtick={0,2,4,6,8,10,12},
    ymin=0, ymax=1.05,
    ytick={0,0.2,0.4,0.6,0.8,1.0},
    xlabel={\textbf{Latency (ms)}},
    ylabel={\textbf{CDF}},
      ylabel style={
        at={(axis description cs:-0.10,0.5)},
        anchor=south,
    },
    grid=major,
    major grid style={dashed, gray!55, line width=0.25pt},
    tick label style={font=\scriptsize},
    label style={font=\scriptsize\bfseries},
    legend style={draw=none, fill=none, font=\scriptsize, at={(0.98,0.02)}, anchor=south east},
    legend cell align={left},
    scaled ticks=false,
    clip=false,
]
\nextgroupplot
\addplot+[const plot mark left, color=myblue, mark=asterisk, mark size=1.1pt, line width=0.75pt] coordinates {
(0.27689123,0.076923077)
(0.30148827,0.15384615)
(0.33084,0.23076923)
(0.36625578,0.30769231)
(0.40952878,0.38461538)
(0.46316751,0.46153846)
(0.61763132,0.53846154)
(0.731838,0.61538462)
(0.88619494,0.69230769)
(1.10199,0.76923077)
(1.9017011,0.84615385)
(2.705982,0.92307692)
(4.18659,1)
};
\addlegendentry{Conf. G}
\addplot+[const plot mark left, color=myorange, mark=asterisk, mark size=1.1pt, line width=0.75pt] coordinates {
(0.37137443,0.076923077)
(0.39597147,0.15384615)
(0.4253232,0.23076923)
(0.46073897,0.30769231)
(0.50401197,0.38461538)
(0.55765071,0.46153846)
(0.71211452,0.53846154)
(0.8263212,0.61538462)
(0.98067813,0.69230769)
(1.1964732,0.76923077)
(1.9961843,0.84615385)
(2.8004652,0.92307692)
(4.2810732,1)
};
\addlegendentry{Conf. F}
\addplot+[const plot mark left, color=mygreen, mark=asterisk, mark size=1.1pt, line width=0.75pt] coordinates {
(0.63159402,0.076923077)
(0.65619105,0.15384615)
(0.68554278,0.23076923)
(0.72095856,0.30769231)
(0.76423156,0.38461538)
(0.8178703,0.46153846)
(0.97233411,0.53846154)
(1.0865408,0.61538462)
(1.2408977,0.69230769)
(1.4566928,0.76923077)
(2.2564039,0.84615385)
(3.0606848,0.92307692)
(4.5412928,1)
};
\addlegendentry{Conf. M}
\addplot+[const plot mark left, color=myred, mark=asterisk, mark size=1.1pt, line width=0.75pt] coordinates {
(1.2918868,0.076923077)
(1.3164838,0.15384615)
(1.3458356,0.23076923)
(1.3812513,0.30769231)
(1.4245243,0.38461538)
(1.4781631,0.46153846)
(1.6326269,0.53846154)
(1.7468336,0.61538462)
(1.9011905,0.69230769)
(2.1169856,0.76923077)
(2.9166967,0.84615385)
(3.7209776,0.92307692)
(5.2015856,1)
};
\addlegendentry{Conf. B}
\node[font=\scriptsize\bfseries] at (rel axis cs:0.5,-0.25) {(a)};
\nextgroupplot
\addplot+[const plot mark left, color=myblue, mark=asterisk, mark size=1.1pt, line width=0.75pt] coordinates {
(0.27689123,0.076923077)
(0.30148827,0.15384615)
(0.33084,0.23076923)
(0.36625578,0.30769231)
(0.40952878,0.38461538)
(0.46316751,0.46153846)
(0.61763132,0.53846154)
(0.731838,0.61538462)
(0.88619494,0.69230769)
(1.10199,0.76923077)
(1.9017011,0.84615385)
(2.705982,0.92307692)
(4.18659,1)
};
\addlegendentry{Conf. G}
\addplot+[const plot mark left, color=myorange, mark=asterisk, mark size=1.1pt, line width=0.75pt] coordinates {
(0.60560075,0.076923077)
(0.63019779,0.15384615)
(0.65954952,0.23076923)
(0.69496529,0.30769231)
(0.73823829,0.38461538)
(0.79187703,0.46153846)
(0.94634084,0.53846154)
(1.0605475,0.61538462)
(1.2149045,0.69230769)
(1.4306995,0.76923077)
(2.2304106,0.84615385)
(3.0346915,0.92307692)
(4.5152995,1)
};
\addlegendentry{Conf. F}
\addplot+[const plot mark left, color=mygreen, mark=asterisk, mark size=1.1pt, line width=0.75pt] coordinates {
(1.5684993,0.076923077)
(1.5930963,0.15384615)
(1.6224481,0.23076923)
(1.6578638,0.30769231)
(1.7011368,0.38461538)
(1.7547756,0.46153846)
(1.9092394,0.53846154)
(2.0234461,0.61538462)
(2.177803,0.69230769)
(2.3935981,0.76923077)
(3.1933092,0.84615385)
(3.9975901,0.92307692)
(5.4781981,1)
};
\addlegendentry{Conf. M}
\addplot+[const plot mark left, color=myred, mark=asterisk, mark size=1.1pt, line width=0.75pt] coordinates {
(3.1656974,0.076923077)
(3.1902944,0.15384615)
(3.2196461,0.23076923)
(3.2550619,0.30769231)
(3.2983349,0.38461538)
(3.3519737,0.46153846)
(3.5064375,0.53846154)
(3.6206441,0.61538462)
(3.7750011,0.69230769)
(3.9907961,0.76923077)
(4.7905072,0.84615385)
(5.5947881,0.92307692)
(7.0753961,1)
};
\addlegendentry{Conf. B}
\node[font=\scriptsize\bfseries] at (rel axis cs:0.5,-0.25) {(b)};
\nextgroupplot[ legend style={draw=none, fill=none, font=\scriptsize, at={(0.98,0.02)}, anchor=south east}]
\addplot+[const plot mark left, color=myblue, mark=asterisk, mark size=1.1pt, line width=0.75pt] coordinates {
(0.27689123,0.076923077)
(0.30148827,0.15384615)
(0.33084,0.23076923)
(0.36625578,0.30769231)
(0.40952878,0.38461538)
(0.46316751,0.46153846)
(0.61763132,0.53846154)
(0.731838,0.61538462)
(0.88619494,0.69230769)
(1.10199,0.76923077)
(1.9017011,0.84615385)
(2.705982,0.92307692)
(4.18659,1)
};
\addlegendentry{Conf. G}
\addplot+[const plot mark left, color=myorange, mark=asterisk, mark size=1.1pt, line width=0.75pt] coordinates {
(1.0886925,0.076923077)
(1.1132896,0.15384615)
(1.1426413,0.23076923)
(1.1780571,0.30769231)
(1.2213301,0.38461538)
(1.2749688,0.46153846)
(1.4294326,0.53846154)
(1.5436393,0.61538462)
(1.6979962,0.69230769)
(1.9137913,0.76923077)
(2.7135024,0.84615385)
(3.5177833,0.92307692)
(4.9983913,1)
};
\addlegendentry{Conf. F}
\addplot+[const plot mark left, color=mygreen, mark=asterisk, mark size=1.1pt, line width=0.75pt] coordinates {
(3.5008665,0.076923077)
(3.5254635,0.15384615)
(3.5548152,0.23076923)
(3.590231,0.30769231)
(3.633504,0.38461538)
(3.6871427,0.46153846)
(3.8416065,0.53846154)
(3.9558132,0.61538462)
(4.1101702,0.69230769)
(4.3259652,0.76923077)
(5.1256763,0.84615385)
(5.9299572,0.92307692)
(7.4105652,1)
};
\addlegendentry{Conf. M}
\addplot+[const plot mark left, color=myred, mark=asterisk, mark size=1.1pt, line width=0.75pt] coordinates {
(7.0304317,0.076923077)
(7.0550287,0.15384615)
(7.0843804,0.23076923)
(7.1197962,0.30769231)
(7.1630692,0.38461538)
(7.2167079,0.46153846)
(7.3711718,0.53846154)
(7.4853784,0.61538462)
(7.6397354,0.69230769)
(7.8555304,0.76923077)
(8.6552415,0.84615385)
(9.4595224,0.92307692)
(10.94013,1)
};
\addlegendentry{Conf. B}
\node[font=\scriptsize\bfseries] at (rel axis cs:0.5,-0.25) {(c)};
\end{groupplot}
\end{tikzpicture}

%% file: Figures/Results/optimization.tex
\begin{figure}
\centering
\begin{tikzpicture}

\begin{axis}[
    name=mainaxis,
    width=0.72\columnwidth,
    height=5.2cm,
    scale only axis,
    xmin=6.5, xmax=22.2,
    ymin=6.4, ymax=14.0,
    xtick={8,10,12,14,16,18,20,22},
    ytick={7,8,9,10,11,12,13},
    axis lines=box,
    grid=both,
    major grid style={gray!55,dashed,line width=0.3pt},
    minor tick num=0,
    xlabel={\textbf{Energy budget, $E_{\max}$ (mJ/bit)}},
    ylabel={\textbf{Energy consumption of}\\\textbf{optimal conf.} (mJ/bit)},
    xlabel style={font=\footnotesize},
    ylabel style={font=\footnotesize, align=center, xshift=0.2em},
    tick label style={font=\footnotesize},
    legend style={
        draw=none,
        fill=none,
        font=\scriptsize,
        at={(0.43,0.58)},
        anchor=west
    },
    legend cell align=left
]

\addplot[
    red,
    dashed,
    thick,
    mark=*,
    mark size=1.5pt,
    mark options={solid}
] coordinates {
    (6.9,6.7) (7.3,6.7) (7.6,6.7) (8.5,6.7) (9.6,6.7)
    (9.85,9.7) (10.25,9.7) (10.65,9.7) (11.10,9.7)
    (12.05,11.85) (12.30,11.85) (12.50,11.85) (12.70,11.85) (13.30,11.85) (13.50,11.85)
    (14.80,13.65) (15.30,13.65) (15.90,13.65) (16.70,13.65) (17.00,13.65) (17.55,13.65)
    (19.00,13.65) (19.35,13.65) (19.50,13.65) (21.40,13.65)
};
\addlegendentry{Energy of selected conf.}

\addlegendimage{blue,dashed,thick,mark=triangle,mark size=2.2pt}
\addlegendentry{Optimised latency}

\end{axis}

\begin{axis}[
    at={(mainaxis.south west)},
    anchor=south west,
    width=0.72\columnwidth,
    height=5.2cm,
    scale only axis,
    xmin=6.5, xmax=22.2,
    ymin=0.22, ymax=1.35,
    xtick=\empty,
    ytick={0.4,0.6,0.8,1.0,1.2},
    axis x line=none,
    axis y line*=right,
    ylabel={\textbf{Optimised latency (ms)}},
    ylabel style={font=\footnotesize, xshift=-1.4em},
    tick label style={font=\footnotesize}
]

\addplot[
    blue,
    dashed,
    thick,
    mark=triangle,
    mark size=2pt,
    mark options={solid},
    forget plot
] coordinates {
    (6.9,1.29) (7.3,1.29) (7.6,1.29) (8.5,1.29) (9.6,1.29)
    (9.85,0.62) (10.25,0.62) (10.65,0.62) (11.10,0.62)
    (12.05,0.37) (12.30,0.37) (12.50,0.37) (12.70,0.37) (13.30,0.37) (13.50,0.37)
    (14.80,0.27) (15.30,0.27) (15.90,0.27) (16.70,0.27) (17.00,0.27) (17.55,0.27)
    (19.00,0.27) (19.35,0.27) (19.50,0.27) (21.40,0.27)
};

\end{axis}

\end{tikzpicture}
\caption{\protect\fontsize{8pt}{9.5pt}\protect\selectfont Energy-aware latency optimization, illustrating energy consumption in the optimal configuration. and the corresponding latency for different values of energy budget $E_{\max}$.}
\label{engVslat}
\end{figure}
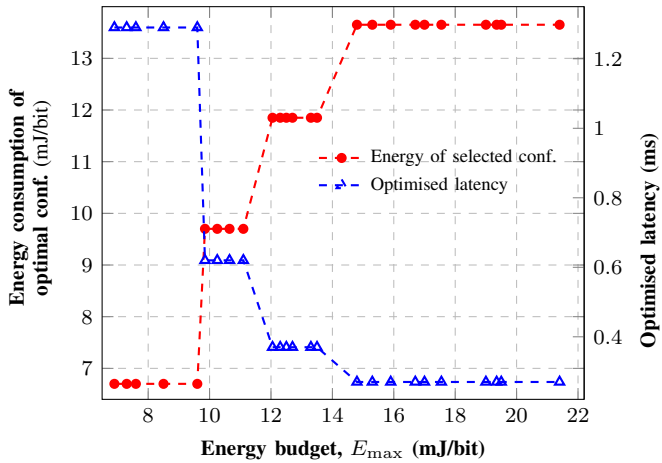